\documentclass[twocolumn,amsmath]{revtex4}
\usepackage{graphicx} 
\usepackage{physics}  
\usepackage{amssymb}  
\usepackage{color}
\usepackage{enumitem}
\usepackage{mathrsfs}

\newcommand{\dket}[1]{\mathinner{|{#1}\rangle \! \rangle}}
\newcommand{\dbra}[1]{\mathinner{\langle \! \langle {#1} |}}
\newcommand{\dbraket}[2]{\mathinner{\langle \! \langle {#1} | {#2} \rangle \! \rangle}}
\newcommand{\dketbra}[2]{\mathinner{|{#1} \rangle \! \rangle \! \langle \! \langle{#2}|}}

\usepackage{amsmath,amsthm}
\newtheorem{prop}{Proposition }

\usepackage{hyperref}
\hypersetup{
colorlinks=true,
linkcolor=blue,
filecolor=magenta,
citecolor=magenta
}

\begin{document}

\title{Time-convolutionless master equation applied to adiabatic elimination}

\author{Masaaki Tokieda}
\email{tokieda.masaaki.4e@kyoto-u.ac.jp}
\affiliation{Department of Chemistry, Graduate School of Science, Kyoto University, Kyoto, Japan}

\author{Angela Riva}
\email{angela.riva@inria.fr}
\affiliation{Laboratoire de Physique de l'Ecole normale supérieure, ENS-PSL, CNRS, Inria, Centre Automatique et Systèmes, Mines Paris, Université PSL, Sorbonne Université, Université Paris Cité, Paris, France}

\begin{abstract}
In open quantum systems theory, reduced models are invaluable for conceptual understanding and computational efficiency.
Adiabatic elimination is a useful model reduction method for systems with separated timescales, where a reduced model is derived by discarding rapidly decaying degrees of freedom.
So far, adiabatic elimination has been formulated using a geometric approach, which provides a versatile and general framework.
This article introduces a reformulation of adiabatic elimination through the framework of the time-convolutionless (TCL) master equation, a widely recognized tool for computing projected time-evolution in open quantum systems.
We show that the TCL master equation formulation yields results equivalent to those obtained from the geometric formulation.
By applying the TCL master equation formulation to typical examples, we demonstrate a practical methodology for performing adiabatic elimination calculation.
This study not only bridges two previously independent approaches, thereby making the adiabatic elimination method accessible to a broader audience, but also enables the analysis of complex cases that are challenging within the geometric formulation.
Additionally, it reveals a geometric interpretation of the TCL master equation formalism.
\end{abstract}

\maketitle

\section{Introduction}

Every quantum system unavoidably interacts with its surrounding environment.
Moreover, the ability to control a quantum system, to manipulate or to read out its state, relies precisely on the possibility to couple it to another system.
These aspects necessitate the modeling of quantum dynamics as those of an open quantum system for a realistic description.
Master equations govern the evolution of states in open quantum systems.
Those equations comprise two components: a Hamiltonian term that generates unitary dynamics, and a non-Hamiltonian term that governs nonunitary, dissipative dynamics. The latter is responsible for the irreversible processes characteristic of open quantum systems.
Prominent examples of master equations include 
the Redfield equation \cite{Redfield57} and the Gorini-Kossakowski-Sudarshan-Lindblad (GKSL) equation \cite{GKS, Lindblad}, which are derived using the Born-Markov approximation \cite{Breuer02} or via stochastic modeling \cite{Budini00,Kiely21}.
These equations capture Markovian dynamics, where memory effects are neglected.
However, we note that non-Markovian dynamics can also be effectively described by incorporating ancilla modes into Markovian master equations \cite{Imamoglu94,Garraway97,Tamascelli18,Xu23}.

In this article, we explore the method of adiabatic elimination, which provides a reduced description for open quantum systems with distinct timescale separations by neglecting rapidly decaying degrees of freedom.
A reduced description is helpful in analyzing systems with large dimensions.
States of open quantum systems are represented by density operators, which entail the number of degrees of freedom that scales quadratically with the system dimension, in contrast to the linear scaling for wave functions. This leads to a significantly higher computational workload for open systems as the system dimension increases.
Adiabatic elimination addresses this challenge by offering a reduced description of the full dynamics, facilitating the investigation of systems that would otherwise be intractable, while still capturing the essential physical behavior in the long-time domain, see e.g. Refs. \cite{Ripoll09,Karabanov15,Halati20}.
Furthermore, a reduced model obtained through adiabatic elimination is crucial in reservoir engineering, where dissipation on the subsystem of interest is engineered by carefully designing the coupling within the open quantum system.
The idea has been employed for diverse applications, including the control of the spin relaxation rate \cite{Bienfait16}, the stabilization of the cat qubit manifold \cite{Mazyar14,Réglade24}, and the observation of a dissipation-induced cross-over in an optical lattice setup \cite{Tomita17}.

There exist various formulations of adiabatic elimination for open quantum systems.
On the one hand, numerous studies have extended techniques originally developed for isolated Hamiltonian systems to address open quantum systems.
These approaches include those that essentially apply the Born-Markov approximation \cite{Cirac92,Wiseman93,Ripoll09,Reiter12,Lesanovsky13,Tomita17,Damanet19,Viana22,Yang22,GB23}, the application of the Laplace transform to the projected master equation \cite{Karabanov15,F-Shapiro20,Saideh20}, and the use of the Schrieffer-Wolff transformation \cite{Kessler12,Burgarth21,Jager22}.
On the other hand, other studies \cite{Macieszczak16,Zanardi16} take advantage of the fact that, in contrast to Hamiltonian systems, the relaxation behavior is inherent in the spectral properties of the generator and trajectories are attracted to a lower-dimensional subspace in the long-time domain.
The idea of leveraging such a geometric picture for adiabatic elimination was further advanced in Ref. \cite{Azouit17}, which is the central focus of our study.
The geometric approach is rooted in classical dynamical system theory, where it was first formulated in the context of singular perturbation theory \cite{Kokot99}, in particular through the Tikhonov approximation theorem \cite{Tik52} and Fenichel's generalization of it \cite{Fenichel79}.
These classical results guarantee, for systems with a distinct timescale separation, the existence of a subspace to which the dynamics converge after an initial fast relaxation phase.
The reduced model of the long-time dynamics on the subspace is thus obtained by discarding the fast relaxation modes.
The geometric approach in Ref. \cite{Azouit17} establishes a  comprehensive framework for calculating higher-order contributions in the timescale separation parameter.
Furthermore, it offers the capability to evaluate the map linking the reduced description to the full density operator of the open quantum system under consideration.
This is in marked contrast to the majority of the other approaches presented above, which primarily concentrate on analyzing the reduced dynamics alone.

This article aims to explore the connection between the geometric approach and the framework of the time-convolutionless (TCL) master equation.
The TCL master equation, developed in seminal works \cite{Tokuyama76,Hashitsume77,Shibata77,Hanggi77}, is a well-established tool in the theory of open quantum systems.
Utilizing a projection onto a subsystem of interest, the framework provides an exact time-local master equation governing the dynamics of the projected state.
Traditionally, the TCL master equation is derived for an isolated system comprising a main system and its environment.
In this study, we instead apply the projection technique to master equations governing open quantum systems with distinct timescales.
As a result, we prove that the TCL master equation framework yields results equivalent to those obtained through the geometric approach to adiabatic elimination, thus establishing a methodological equivalence between these two approaches.

This study bridges the conceptual gap between the two approaches.
It consolidates model reduction techniques from classical perturbation theory for dynamical systems on the one hand, and projection methods for deriving reduced master equations on the other, into a unified theoretical framework.
Furthermore, by reformulating adiabatic elimination within the TCL master equation framework that is widely recognized in the open quantum systems community, this study makes the adiabatic elimination method more accessible to a broader audience.

In addition to the aforementioned aspects, this study uncovers insights into both the geometric approach and the TCL master equation formalism. 
The projection techniques in the TCL master equation framework turn out to provide a notable simplification for scenarios previously challenging to address using only the geometric approach.
Noteworthy examples include the occurrence of rapid unitary dynamics in the absence of perturbation, as well as instances where short-time dynamics preceding the decay of fast degrees of freedom influence the long-term dynamics.
While the geometric approach requires new formulations to handle these scenarios \cite{Angela24,FM23}, the TCL master equation framework offers a consistent and straightforward solution, as demonstrated through an example in this article.
Furthermore, the proof of the equivalence provides insights into the TCL master equation formalism.
We show that a superoperator arising in the derivation of the TCL master equation corresponds to the projection onto the lower-dimensional subspace to which the full dynamics converge after an initial fast relaxation phase.
Therefore, this study offers a geometric interpretation of the TCL master equation formalism, which has traditionally been formulated solely through analytical means. 

The rest of this article is organized as follows.
In Sec. \ref{AE}, we review the geometric formulation of adiabatic elimination.
In Sec. \ref{TCL}, we reformulate adiabatic elimation using the TCL master equation framework.
We start by reviewing the TCL master equation in Sec. \ref{TCL_review}.
We then show in Sec. \ref{TCL_ad.el.} that the TCL master equation formulation provides results consistent with the geometric formulation. 
Section \ref{Demo} is dedicated to practical demonstrations of our formulation.
We consider a three level system to numerically verify the consistency in Sec. \ref{Demo_prop}.
We then analyze the classic example of bipartite systems in Sec. \ref{Demo_bipartite}.
Finally, a summary and concluding remarks are presented in Sec. \ref{summary}.

\section{Geometric approach to adiabatic elimination}
\label{AE}

In this section, we present the geometric approach to adiabatic elimination, which was originally formulated for bipartite GKSL systems in Ref. \cite{Azouit17} and later extended to more general settings in Ref. \cite{FM23}.
We consider a master equation,
\begin{equation}
    \frac{d}{dt} \rho(t) = ( \mathcal{L}_0 + \epsilon \mathcal{L}_1 ) \rho(t) \equiv \mathcal{L} \rho(t),
    \label{eq:AE_master}
\end{equation}
where $\rho (t)$ is the density operator and $\epsilon$ is a small nonnegative parameter.
The superoperator $\mathcal{L}_0$ is assumed to be diagonalizable in this article.
Henceforth, we denote the identity superoperator as $\mathcal{I}$ and the identity operator or matrix as $I$.

\begin{figure}[t]
  \includegraphics[keepaspectratio, scale=0.2]{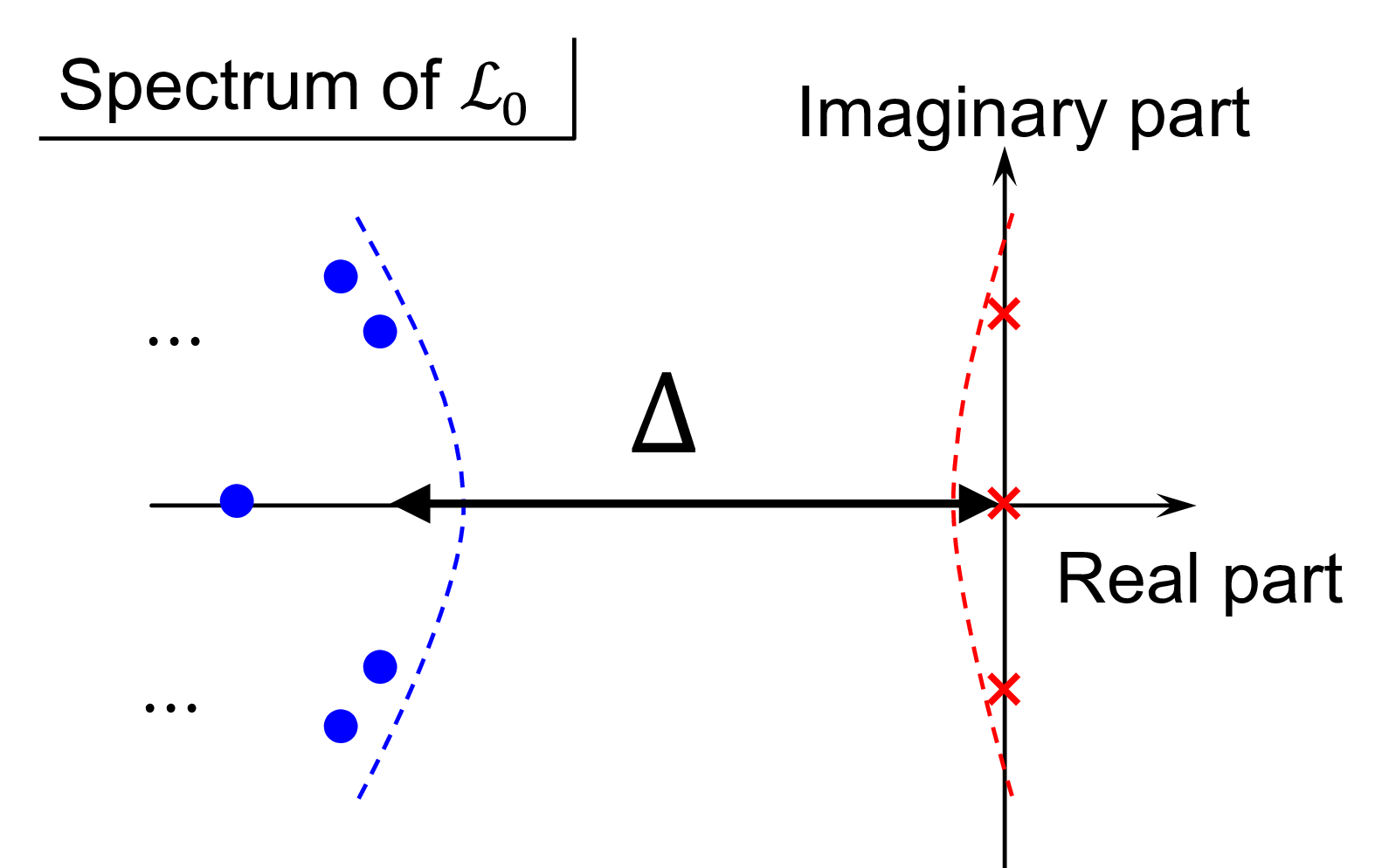}
  \caption{
  Typical spectrum of $\mathcal{L}_0$ [see Eq. (\ref{eq:AE_master})] in adiabatic elimination.
  The presence of a gap $\Delta$ is assumed between surviving modes proximate to the imaginary axis (represented by the red crosses) and fast relaxation modes (represented by the blue circles).
  }
  \label{fig:AE_gap}
\end{figure}

We assume the presence of a gap $\Delta$ in the spectrum of $\mathcal{L}_0$, as illustrated in Fig. \ref{fig:AE_gap}.
When $\epsilon = 0$, modes represented by the blue circles in Fig. \ref{fig:AE_gap} decay in the time regime $t \gg \Delta^{-1}$.
Consequently, modes close to the imaginary axis (i.e., the vertical line corresponding to the zero real part), represented by the red crosses in the same figure, are the sole modes to survive asymptotically.
For those surviving modes, we assume that the eigenvalue problem of $\mathcal{L}_0$ is solvable.
The eigenvalue problem of $\mathcal{L}_0$ reads
\begin{equation}
    (\hat{\mathcal{L}}_0 - \lambda_i \hat{\mathcal{I}}) \dket{r_i} = 0, \ \ \dbra{l_i} (\hat{\mathcal{L}}_0 - \lambda_i \hat{\mathcal{I}}) = 0,
    \label{eq:AE_eig.L0}
\end{equation}
where we have introduced $\dket{A}$ as a column vector representation of an operator $A$ and $\dbra{A}$ as its Hermitian conjugate such that the inner product is given by the Hilbert-Schmidt inner product $\dbraket{A}{B} = {\rm tr} (A^\dagger B)$ with ${\rm tr}$ the trace operation.
In the vectorized representation, superoperators are represented as matrices, which we distinguish by attaching a hat ( $\hat{}$ ).
In the above equation, $\lambda_i$ is the eigenvalue and $\dket{r_i}$ and $\dbra{l_i}$ are the right and left eigenvectors, respectively, which are normalized as $\dbraket{l_i}{r_j} = \delta_{ij}$ with $\delta_{ij}$ the Kronecker delta.

In what follows, we use the subscript $i$ for arbitrary modes, $s$ (surviving) for the surviving modes, and $f$ (fast relaxation) for the other modes.
With these notations, the gap $\Delta$ can be defined as $\Delta = \min_f {\rm Re}( - \lambda_f )$ with ${\rm Re}$ denoting the real part.
As illustrated in Fig. \ref{fig:AE_gap}, we assume ${\rm Re} \lambda_s = 0 \ (\forall s)$.
On the other hand, ${\rm Im} \lambda_s$, with ${\rm Im}$ denoting the imaginary part, need not be zero. 
In other words, the surviving modes allow rapid unitary dynamics, when ${\rm Im} \lambda_s$ has a substantial magnitude.
Hence, the formulation in this article can be applied to a broader class of problems compared to those addressed in Ref. \cite{FM23}, where $\lambda_s = 0 \ (\forall s)$ were assumed.

\begin{figure}[t]
  \includegraphics[keepaspectratio, scale=0.25]{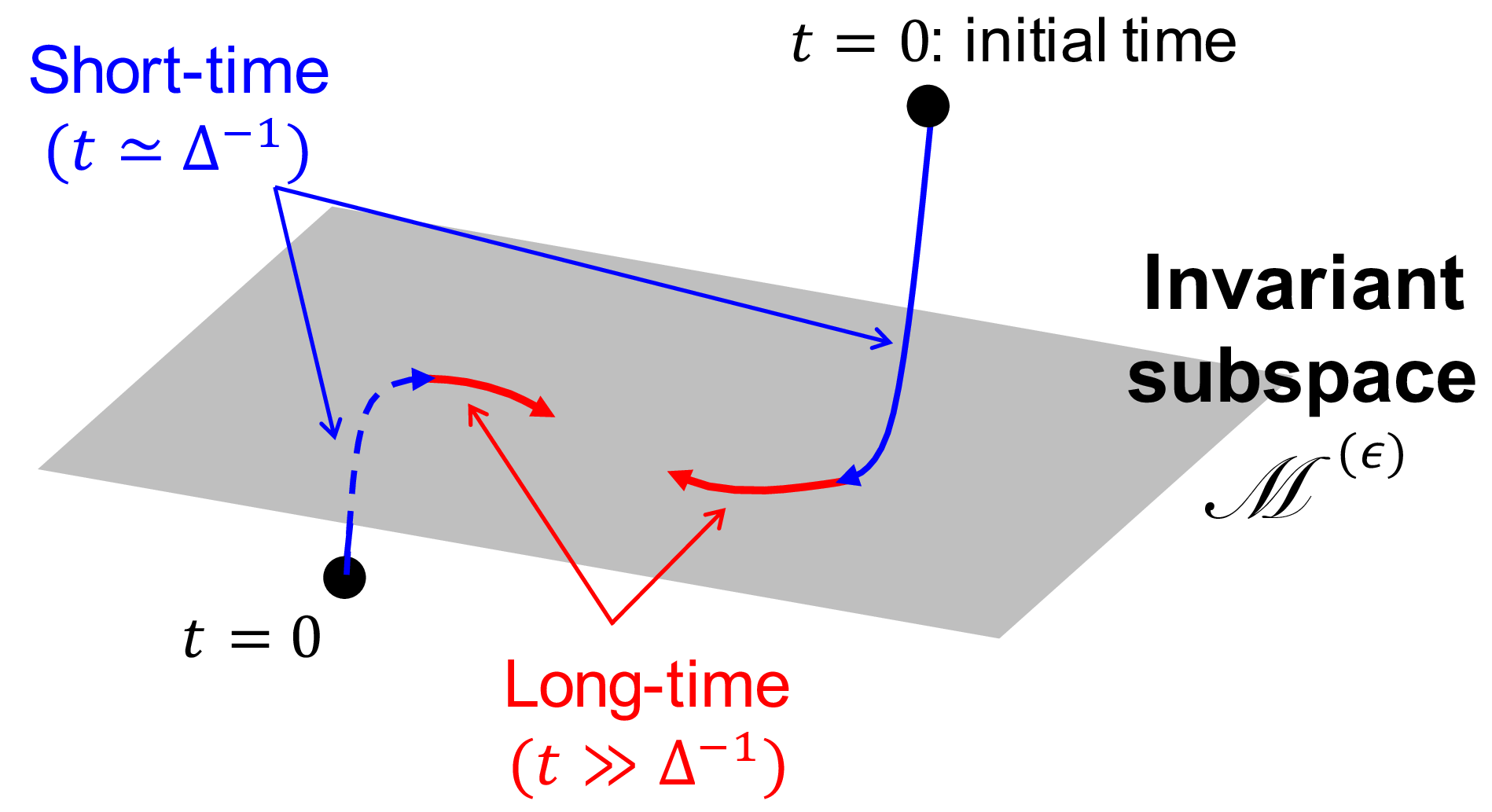}
  \caption{
  Schematic illustration of the total state evolution in adiabatic elimination.
  The gray plane represents the invariant subspace $\mathscr{M}^{(\epsilon)}$, which is preserved by the operation of $\mathcal{L}$.
  Starting from an arbitrary initial state, in the short-time regime ($t \approx \Delta^{-1}$), the state is rapidly attracted to $\mathscr{M}^{(\epsilon)}$.
  In the long-time regime ($t \gg \Delta^{-1}$), the state is restricted within $\mathscr{M}^{(\epsilon)}$.
  }
  \label{fig:AE_trajectory}
\end{figure}

Adiabatic elimination achieves a reduction of the model space dimension by discarding fast relaxation modes in order to describe the long-time behavior.
In the presence of a gap $\Delta$ in the spectrum of $\mathcal{L}_0$, the dynamics generated by $\mathcal{L}_0$ typically begin with a rapid decaying phase described by the fast relaxation modes (the blue circles in Fig. \ref{fig:AE_gap}).
This phase is subsequently followed by an evolution dominated by the surviving modes (the red crosses in Fig. \ref{fig:AE_gap}).
At times $t$ such that $\exp(- t \Delta)$ is negligible, hence, the dynamics are closed within a lower-dimensional subspace spanned by the surviving modes.
This subspace is called invariant subspace, since it is preserved by the operation of $\mathcal{L}_0$.
As long as the matrix elements of $\epsilon \mathcal{L}_1$ with respect to the eigenvectors of $\mathcal{L}_0$ are sufficiently small compared to the gap $\Delta$, this picture still holds true with nonzero $\epsilon$ \cite{Fenichel79}.
That is, the state first experiences rapid relaxation toward the invariant subspace (hereafter denoted by $\mathscr{M}^{(\epsilon)}$), which is invariant under the operation of $\mathcal{L}$, and then exhibits a slower relaxation dynamics in it, as illustrated in Fig. \ref{fig:AE_trajectory}.
Given this picture, the goal of adiabatic elimination is to describe the dynamics in $\mathscr{M}^{(\epsilon)}$, which requires fewer degrees of freedom compared to the original problem Eq. (\ref{eq:AE_master}).

To formulate the aforementioned picture, we parametrize the degrees of freedom in $\mathscr{M}^{(\epsilon)}$ and denote them by $\vec{x}$.
We then calculate two maps that are assumed to be linear and time-independent.
One, denoted by $\mathcal{K}^{(\epsilon)}$, relates the parametrization to the solution to the master equation (\ref{eq:AE_master}), $\dket{\rho(t)} = \mathcal{K}^{(\epsilon)} \vec{x}(t)$.
The other, denoted by $\mathcal{F}^{(\epsilon)}$, describes the time evolution of the parameters $\vec{x}$, $(d/dt) \vec{x}(t) = \mathcal{F}^{(\epsilon)} \vec{x}(t)$.
When $\epsilon = 0$, a state in $\mathscr{M}^{(\epsilon = 0)}$ can be represented by a linear combination of $\{ \dket{r_s} \}$.
Hence, we fix $\mathcal{K}^{(\epsilon = 0)}$ as
\begin{equation}
    \mathcal{K}^{(\epsilon = 0)} \vec{x} = \sum_s x_s \dket{r_s}.
    \label{eq:AE_K0}
\end{equation}

Inserting the definitions of the maps into Eq. (\ref{eq:AE_master}), we obtain the invariance condition,
\begin{equation}
    \mathcal{K}^{(\epsilon)} \mathcal{F}^{(\epsilon)} = \hat{\mathcal{L}} \mathcal{K}^{(\epsilon)}.
    \label{eq:AE_inv.cond.}
\end{equation}
The maps $\mathcal{K}^{(\epsilon)}$ and $\mathcal{F}^{(\epsilon)}$ can be determined from this equation.
For instance, the authors of Ref. \cite{Alain20} found analytic solutions to the invariance condition for a dispersively coupled qubit-qubit system, which was later extended to qubit-qudit ($d$-dimensional system with any positive integer $d$) systems in Ref. \cite{TESR23}.
In most examples, however, it is difficult to find the maps $\mathcal{K}^{(\epsilon)}$ and $\mathcal{F}^{(\epsilon)}$ satisfying Eq. (\ref{eq:AE_inv.cond.}) analytically.
In such cases, one can perform the asymptotic expansion of those maps with respect to $\epsilon$ in order to find a solution approximately. We provide the details in Appendix \ref{app:inv.cond.}.

Note that the solution to Eq. (\ref{eq:AE_inv.cond.}) is not unique.
To see this, let $\mathcal{T}$ be an invertible map (not necessarily a unitary map) of the same dimension as $\mathcal{F}^{(\epsilon)}$.
If $\mathcal{K}^{(\epsilon)}$ and $\mathcal{F}^{(\epsilon)}$ satisfy Eq. (\ref{eq:AE_inv.cond.}), then the maps defined by $\mathcal{K}^{(\epsilon)} \mathcal{T} \equiv \bar{\mathcal{K}}^{(\epsilon)}$ and $\mathcal{T}^{-1} \mathcal{F}^{(\epsilon)} \mathcal{T} \equiv \bar{\mathcal{F}}^{(\epsilon)}$ also satisfy the invariance condition ($\bar{\mathcal{K}}^{(\epsilon)}$ is consistent with Eq. (\ref{eq:AE_K0}) assuming $\lim_{\epsilon \to 0} \mathcal{T} = \mathcal{I}$).
Given the definitions $\dket{\rho(t)} = \mathcal{K}^{(\epsilon)} \vec{x}(t)$ and $(d/dt) \vec{x}(t) = \mathcal{F}^{(\epsilon)} \vec{x}(t)$, we can show that $\dket{\rho(t)} = \bar{\mathcal{K}}^{(\epsilon)} \vec{y}(t)$ and $(d/dt) \vec{y}(t) = \bar{\mathcal{F}}^{(\epsilon)} \vec{y} (t)$ with $\vec{y}(t) = \mathcal{T}^{-1} \vec{x}(t)$.
These identities indicate that the solutions $\bar{\mathcal{K}}^{(\epsilon)}$ and $\bar{\mathcal{F}}^{(\epsilon)}$ are the maps for the different parametrization $\vec{y}(t)$.
Thus, the gauge degree of freedom is associated with the nonuniqueness of the parametrization.
In what follows, we consider only a gauge choice such that, with Eq. (\ref{eq:AE_K0}),
\begin{equation}
    \dbra{l_s} ( \mathcal{K}^{(\epsilon)} - \mathcal{K}^{(\epsilon = 0)} ) \vec{x} = 0,
    \label{eq:AE_gauge}
\end{equation}
for any $s$ and $\vec{x}$.
This choice is equivalent to considering the parametrization given by $x_s(t) = \dbraket{l_s}{\rho(t)}$.
As we discuss later, this is a natural representation in practice [see Eq. (\ref{eq:Demo_rhoB})].

\section{Adiabatic elimination through the TCL master equation framework}
\label{TCL}

In this section, we discuss that adiabatic elimination presented in Sec. \ref{AE} can be equivalently formulated using the TCL master equation.
We first review the TCL master equation in Sec. \ref{TCL_review}.
We then discuss the equivalence in Sec. \ref{TCL_ad.el.}.

\subsection{TCL master equation}
\label{TCL_review}

We begin by reviewing the TCL master equation, which has been widely applied to analyze effects of the environment on the relevant system dynamics, see e.g. Refs. \cite{Lengers20,Lupke20,Gulacsi23} for recent contributions.
Although the method has primarily been used to compute the reduced dynamics, the total density operator can also be extracted as shown in Ref. \cite{Trushechkin19} and below.

We follow the derivation presented in Refs. \cite{Shibata77,Breuer02}.
Let $\mathcal{P}$ be a projection superoperator satisfying $\mathcal{P}^2 = \mathcal{P}$.
The subsequent derivation is independent of the explicit form of $\mathcal{P}$.
The specification of its precise form is thus deferred until the discussion on the equivalence to adiabatic elimination in Sec. \ref{TCL_ad.el.}.
The superoperator projecting onto the complementary subspace is given by $\mathcal{Q} = \mathcal{I} - \mathcal{P}$.
From its construction, it is evident that $\mathcal{Q}$ satisfies $\mathcal{Q}^2 = \mathcal{Q}$ and $\mathcal{P} \mathcal{Q} = \mathcal{Q} \mathcal{P} = 0$

The TCL master equation is an evolution equation for $\mathcal{P} \rho(t)$.
From Eq. (\ref{eq:AE_master}), the coupled equations for $\mathcal{P} \rho(t)$ and $\mathcal{Q} \rho(t)$ can be derived as
\begin{equation*}
    \begin{gathered}
        \frac{d}{dt} \mathcal{P} \rho(t) = \mathcal{P} \mathcal{L} (\mathcal{P} + \mathcal{Q}) \rho(t), \\
        \frac{d}{dt} \mathcal{Q} \rho(t) = \mathcal{Q} \mathcal{L} (\mathcal{P} + \mathcal{Q}) \rho(t).        
    \end{gathered}
\end{equation*}
Solving the equation for $\mathcal{Q} \rho(t)$ formally, we obtain
\begin{equation}
    \begin{gathered}
        \mathcal{Q} \rho(t) = e^{\mathcal{Q} \mathcal{L} \mathcal{Q} t } \mathcal{Q} \rho(0) \\
        + \int_{0}^t ds \, e^{\mathcal{Q} \mathcal{L} \mathcal{Q} (t-s) } \mathcal{Q} \mathcal{L} \mathcal{P} \rho(s),
    \end{gathered}
    \label{eq:TCL_Q.rho.conv}
\end{equation}
where the initial time is set to 0.
Inserting Eq. (\ref{eq:TCL_Q.rho.conv}) into the equation for $\mathcal{P} \rho(t)$ yields the so called Nakajima-Zwanzig equation \cite{Nakajima58,Zwanzig60};
\begin{equation}
    \begin{gathered}
        \frac{d}{dt} \mathcal{P} \rho(t) = \mathcal{P} \mathcal{L} e^{\mathcal{Q} \mathcal{L} \mathcal{Q} t } \mathcal{Q} \rho(0) \\
        + \mathcal{P} \mathcal{L} \mathcal{P} \rho(t) + \int_{0}^t ds \, \mathcal{P} \mathcal{L} \mathcal{Q} e^{\mathcal{Q} \mathcal{L} \mathcal{Q} (t-s) } \mathcal{Q} \mathcal{L} \mathcal{P} \rho(s).
    \end{gathered}
    \label{eq:TCL_NZ}
\end{equation}
This equation contains the time-convolution integral (see the second term in the second line), which comes from the second line in Eq. (\ref{eq:TCL_Q.rho.conv}).
Hence, the time-convolution integral can be removed by replacing $\rho(s)$ in the integral with $\rho(t)$.
To this end, we evolve $\rho(t)$ backward in time to $\rho(s)$,
\begin{equation*}
    \rho(s) = e^{\mathcal{L} (s-t)} \rho(t).
\end{equation*}
Inserting this into Eq. (\ref{eq:TCL_Q.rho.conv}) then yields
\begin{equation*}
    [\mathcal{I} - \Sigma(t)] \mathcal{Q} \rho(t) =  e^{\mathcal{Q} \mathcal{L} \mathcal{Q} t } \mathcal{Q} \rho(0) + \Sigma(t) \mathcal{P} \rho (t),
\end{equation*}
with 
\begin{equation}
    \Sigma(t) = \int_{0}^{t} d\tau \, e^{\mathcal{Q} \mathcal{L} \mathcal{Q} \tau } \mathcal{Q} \mathcal{L} \mathcal{P} e^{- \mathcal{L} \tau}.
    \label{eq:TCL_Sigma_general}
\end{equation}

We examine two distinct scenarios in this article. 
The first scenario is where $[\mathcal{P}, \mathcal{L}_0] = 0$.
We can then replace $\mathcal{Q} \mathcal{L} \mathcal{P}$ by $\mathcal{Q} \mathcal{L}_1 \mathcal{P}$ due to $\mathcal{Q} \mathcal{P} = 0$:
\begin{equation}
    \Sigma(t) = \epsilon \int_{0}^{t} d\tau \, e^{\mathcal{Q} \mathcal{L} \mathcal{Q} \tau } \mathcal{Q} \mathcal{L}_1 \mathcal{P} e^{- \mathcal{L} \tau}.
    \label{eq:TCL_Sigma}
\end{equation}
We assume the existence of $[\mathcal{I} - \Sigma(t)]^{-1}$, which is valid for small $\epsilon$.
By inverting $[\mathcal{I} - \Sigma(t)]$ and using $[\mathcal{I} - \Sigma(t)]^{-1} \Sigma(t) = [\mathcal{I} - \Sigma(t)]^{-1} - \mathcal{I}$, we obtain an expression of $\mathcal{Q} \rho(t)$ without the time-convolution integral,
\begin{equation*}
    \mathcal{Q} \rho(t) = \mathcal{J}(t) \mathcal{Q} \rho(0) + \Big\{ [\mathcal{I} - \Sigma(t)]^{-1} - \mathcal{I} \Big\} \mathcal{P} \rho(t),
\end{equation*}
where 
\begin{equation}
    \mathcal{J}(t) = [\mathcal{I} - \Sigma(t)]^{-1} e^{\mathcal{Q} \mathcal{L} \mathcal{Q} t} \mathcal{Q}.
    \label{eq:TCL_Jt}
\end{equation}
This equation indicates
\begin{equation}
    \rho(t) = \mathcal{J}(t) \mathcal{Q} \rho(0) + \mathcal{P}(t) \rho(t),
    \label{eq:TCL_Q.rho.}
\end{equation}
with
\begin{equation}
    \mathcal{P}(t) = [\mathcal{I} - \Sigma(t)]^{-1} \mathcal{P}.
    \label{eq:TCL_Pt}
\end{equation}
We can show $[ \mathcal{P}(t) ]^2 = \mathcal{P}(t)$ from $\mathcal{P} \Sigma(t) = 0$. Hence, $\mathcal{P}(t)$ is also a projection.
Inserting Eq. (\ref{eq:TCL_Q.rho.}) into the right hand side of $(d/dt) \mathcal{P} \rho(t) = \mathcal{P} \mathcal{L} \rho(t)$, we obtain the TCL master equation
\begin{equation}
    \frac{d}{dt} \mathcal{P} \rho (t) = \mathcal{P} \mathcal{L} \mathcal{J}(t) \mathcal{Q} \rho(0) + \mathcal{P} \mathcal{L} \mathcal{P}(t) \rho(t).
    \label{eq:TCL_TCL}
\end{equation}
Note that, in Eqs. (\ref{eq:TCL_Q.rho.}) and (\ref{eq:TCL_TCL}), the terms involving $\mathcal{J}(t)$ explicitly depend on the initial condition in the complementary subspace $\mathcal{Q} \rho(0)$.

The second scenario is where $\Sigma(t) = 0$.
In this case, we obtain $\mathcal{Q} \rho(t) = \exp( \mathcal{Q} \mathcal{L} \mathcal{Q} t) \mathcal{Q} \rho(0)$, which yields
\begin{equation}
    \rho(t) = e^{\mathcal{Q} \mathcal{L} \mathcal{Q} t} \mathcal{Q} \rho(0) + \mathcal{P} \rho(t),
    \label{eq:TCL_Q.rho.2nd}
\end{equation}
and, by inserting it into the right hand side of $(d/dt) \mathcal{P} \rho(t) = \mathcal{P} \mathcal{L} \rho(t)$,
\begin{equation}
    \frac{d}{dt} \mathcal{P} \rho (t) = \mathcal{P} \mathcal{L} e^{\mathcal{Q} \mathcal{L} \mathcal{Q} t} \mathcal{Q} \rho(0) + \mathcal{P} \mathcal{L} \mathcal{P} \rho(t).
    \label{eq:TCL_TCL.2nd}
\end{equation}

\subsection{Maps $\mathcal{K}^{(\epsilon)}$ and $\mathcal{F}^{(\epsilon)}$ from the TCL master equation}
\label{TCL_ad.el.}

We here show that the TCL master equation framework developed above provides an alternative formulation of adiabatic elimination.
For this purpose, the main task is to show that the maps $\mathcal{K}^{(\epsilon)}$ and $\mathcal{F}^{(\epsilon)}$ can be computed using quantities in the TCL master equation framework since the goal of adiabatic elimination is to evaluate those maps.
In the geometric formulation, those maps are obtained as a solution to the invariance condition (\ref{eq:AE_inv.cond.}).
The invariance condition is solved under two conditions: the boundary condition (\ref{eq:AE_K0}) and the gauge fixing condition (\ref{eq:AE_gauge}).
These conditions determine the parametrization of $\mathscr{M}^{(\epsilon)}$.
Hence, we need to prove the existence of maps corresponding to $\mathcal{K}^{(\epsilon)}$ and $\mathcal{F}^{(\epsilon)}$ within the framework of the TCL master equation, which we denote as $\mathcal{K}_{\rm TCL}^{(\epsilon)}$ and $\mathcal{F}_{\rm TCL}^{(\epsilon)}$, respectively,
such that $\mathcal{K}_{\rm TCL}^{(\epsilon)}$ satisfies the boundary condition and the gauge fixing condition and that $\mathcal{K}_{\rm TCL}^{(\epsilon)}$ and $\mathcal{F}_{\rm TCL}^{(\epsilon)}$ satisfy the invariance condition.

\subsubsection*{Initial states outside the invariant subspace $\mathscr{M}^{(\epsilon)}$}

We first consider scenarios with $\rho(0) \not\in \mathscr{M}^{(\epsilon)}$, that is, the initial state $\rho(0)$ is outside the invariant subspace with $\epsilon > 0$.
The geometric formulation, which assumes that the initial state is in $\mathscr{M}^{(\epsilon)}$, provides an approximate description of the long-time behavior.
To provide physical context, one example of this scenario is quench dynamics, where the system evolves under $\mathcal{L}_0$ for $t < 0$, relaxing to the corresponding invariant subspace such that $\rho(0) \in \mathscr{M}^{(\epsilon = 0)}$, and the perturbation $\epsilon \mathcal{L}_1$ is abruptly introduced at $t = 0$.

In this case, we adopt $\mathcal{P} = \mathcal{P}_{\rm inv}$ with 
\begin{equation}
    \hat{\mathcal{P}}_{\rm inv} = \sum_s \dketbra{r_s}{l_s},
    \label{eq:TCL_P.def}
\end{equation}
the operation of which is given by $\mathcal{P}_{\rm inv} \rho = \sum_s {\rm tr}(l_s^\dagger \rho) r_s$.
This is a projection onto $\mathscr{M}^{(\epsilon = 0)}$.
In fact, the orthonormality condition of the eigenvectors ensures the property $\mathcal{P}_{\rm inv}^2 = \mathcal{P}_{\rm inv}$, which corroborates its nature as a projection.
Furthermore, we have the commutation relation $[\mathcal{L}_0, \mathcal{P}_{\rm inv}] = 0$, as expected from the fact that $\mathcal{P}_{\rm inv}$ is a projection onto eigenspaces of $\mathcal{L}_0$.
This implies that the corresponding $\Sigma(t)$ defined in Eq. (\ref{eq:TCL_Sigma_general}), which we denote as $\Sigma_{\rm inv}^{(\epsilon)}(t)$, is given by Eq. (\ref{eq:TCL_Sigma}) with $\mathcal{P} = \mathcal{P}_{\rm inv}$ and $\mathcal{Q} = \mathcal{I} - \mathcal{P}_{\rm inv} \equiv \mathcal{Q}_{\rm inv}$.
The corresponding $\mathcal{J}(t)$ [Eq. (\ref{eq:TCL_Jt})] and $\mathcal{P}(t)$ [Eq. (\ref{eq:TCL_Pt})], which we denote as $\mathcal{J}_{\rm inv}^{(\epsilon)}(t)$ and $\mathcal{P}_{\rm inv}^{(\epsilon)}(t)$, respectively, are given by
\begin{equation}
    \mathcal{J}_{\rm inv}^{(\epsilon)}(t) = [\mathcal{I} - \Sigma_{\rm inv}^{(\epsilon)}(t)]^{-1} e^{ \mathcal{Q}_{\rm inv} \mathcal{L} \mathcal{Q}_{\rm inv} t } \mathcal{Q}_{\rm inv},
    \label{eq:TCL_J.inv.eps}
\end{equation}
and 
\begin{equation}
    \mathcal{P}_{\rm inv}^{(\epsilon)} (t) = [\mathcal{I} - \Sigma_{\rm inv}^{(\epsilon)}(t)]^{-1} \mathcal{P}_{\rm inv}.
    \label{eq:TCL_P.inv.eps}
\end{equation}
With these notations, Eqs. (\ref{eq:TCL_Q.rho.}) and (\ref{eq:TCL_TCL}) read in the vector representation as
\begin{equation}
\begin{gathered}
    \dket{\rho(t)} = \hat{\mathcal{J}}_{\rm inv}^{(\epsilon)}(t)  \dket{\mathcal{Q}_{\rm inv} \rho(0)} + \hat{\mathcal{P}}_{\rm inv}^{(\epsilon)}(t) \dket{\rho(t)}, \\
    \frac{d}{dt} \hat{\mathcal{P}}_{\rm inv} \dket{\rho (t)} = \hat{\mathcal{P}}_{\rm inv} \hat{\mathcal{L}} \hat{\mathcal{J}}_{\rm inv}^{(\epsilon)}(t) \dket{\mathcal{Q}_{\rm inv} \rho(0)} \\
    + \hat{\mathcal{P}}_{\rm inv} \hat{\mathcal{L}} \hat{\mathcal{P}}_{\rm inv}^{(\epsilon)}(t) \dket{\rho(t)}.
\end{gathered}
\label{eq:TCL_full.rho}
\end{equation}

Now we incorporate the gauge fixing condition (\ref{eq:AE_gauge}).
As discussed below Eq. (\ref{eq:AE_gauge}), it is equivalent to $x_s(t) = \dbraket{l_s}{\rho(t)}$ for any surviving modes $s$.
For a more compact description, let us introduce 
\begin{equation}
    \chi_R = \left[ \dket{r_{s=1}} \, \dket{r_{s=2}} \, \cdots \right],  \ \ \chi_L^\dagger =
    \left[
    \begin{matrix}
        \dbra{l_{s=1}} \\
        \dbra{l_{s=2}} \\
        \vdots
    \end{matrix}
    \right].
    \label{eq:TCL_chiRchiL}
\end{equation}
We find $\chi_R \chi_L^\dagger = \hat{\mathcal{P}}_{\rm inv}$, $\chi_L^\dagger \chi_R = I$, and $\vec{x} (t) = \chi_L^\dagger \dket{\rho(t)}$.
Inserting these into Eqs. (\ref{eq:TCL_full.rho}), the equations can be expressed with $\vec{x}(t)$ as
\begin{equation}
\begin{gathered}
    \dket{\rho(t)} = \hat{\mathcal{J}}_{\rm inv}^{(\epsilon)}(t) \dket{\mathcal{Q}_{\rm inv} \rho(0)} + \hat{\mathcal{P}}_{\rm inv}^{(\epsilon)}(t) \chi_R \vec{x}(t) \\
    \frac{d}{dt} \vec{x}(t) = \chi_L^\dagger  \hat{\mathcal{L}} \hat{\mathcal{J}}_{\rm inv}^{(\epsilon)}(t) \dket{\mathcal{Q}_{\rm inv} \rho(0)} \\
    + \chi_L^\dagger \hat{\mathcal{L}} \hat{\mathcal{P}}_{\rm inv}^{(\epsilon)}(t) \chi_R \, \vec{x}(t).
\end{gathered}
\label{eq:TCL_full.x}
\end{equation}

Let us denote $\mathcal{P}_{\rm inv}^{(\epsilon)} (t)$ in the asymptotic time limit as $\mathcal{P}_{\rm inv}^{(\epsilon)} = \lim_{t \to \infty} \mathcal{P}_{\rm inv}^{(\epsilon)} (t)$.
To proceed, we note the following two properties of $\mathcal{J}_{\rm inv}^{(\epsilon)}(t)$ and $\mathcal{P}_{\rm inv}^{(\epsilon)}(t) - \mathcal{P}_{\rm inv}^{(\epsilon)}$.
First, for small $\epsilon$, these superoperators decay exponentially in time;
\begin{prop}
    It follows that
    \begin{equation}
        \| \hat{\mathcal{P}}_{\rm inv}^{(\epsilon)} (t) - \hat{\mathcal{P}}_{\rm inv}^{(\epsilon)} \| = O \left( (\epsilon / \Delta) e^{- t \Delta} \right),
        \label{eq:TCL_P.relax}
    \end{equation}
    and
    \begin{equation}
        \| \hat{\mathcal{J}}_{\rm inv}^{(\epsilon)} (t) \| = O \left( e^{- t \Delta} \right),
        \label{eq:TCL_J.relax}
    \end{equation}
    as $\epsilon \to 0$, where $\| \bullet \|$ denotes a matrix norm.
    \label{prop:relax}
\end{prop}
\noindent For the second property, recall that $\mathcal{P}_{\rm inv}^{(\epsilon)} (t)$ is a projection as shown below Eq. (\ref{eq:TCL_Pt}).
In this regard, the projection in the asymptotic time limit $\mathcal{P}_{\rm inv}^{(\epsilon)}$ has a distinct geometric interpretation as follows.
\begin{prop}
    The image of the projection $\mathcal{P}_{\rm inv}^{(\epsilon)}$ is $\mathscr{M}^{(\epsilon)}$, that is, if $\dket{r_s^{(\epsilon)}}$ is the right eigenvector of $\mathcal{L}$ associated with $\dket{r_s}$ in the limit $\epsilon \to 0$, the projection $\mathcal{P}_{\rm inv}^{(\epsilon)}$ can be expressed as
    \begin{equation}
        \hat{\mathcal{P}}_{\rm inv}^{(\epsilon)} = \sum_{s s'} \dket{r_{s}^{(\epsilon)}} [N^{-1}]_{s s'} \dbra{l_{s'}},
        \label{eq:TCL_P.tinf}
    \end{equation}
    with $N_{s s'} = \dbraket{l_{s}}{r_{s'}^{(\epsilon)}}$. In particular, it satisfies 
    \begin{equation}
        (\mathcal{I} - \mathcal{P}_{\rm inv}^{(\epsilon)}) \mathcal{L} \mathcal{P}_{\rm inv}^{(\epsilon)} = 0.
        \label{eq:TCL_proj.inv.}
    \end{equation}
    \label{prop:proj}
\end{prop}
\noindent
The proofs of these propositions are provided in Appendix \ref{app:proof}.
We consider finite-dimensional systems.
Assuming small $\epsilon$ to maintain the gap structure of the generator $\mathcal{L} = \mathcal{L}_0 + \epsilon \mathcal{L}_1$, we employ the spectral decomposition of $\mathcal{L}$ to analyze the time-dependence of $\mathcal{P}_{\rm inv}^{(\epsilon)}(t)$ and $\mathcal{J}_{\rm inv}^{(\epsilon)}(t)$.
The gap ensures decay of the time-dependent terms, thereby establishing Propositions \ref{prop:relax} and \ref{prop:proj}.
Notably, the proof does not rely on perturbation expansions and Eq. (\ref{eq:TCL_proj.inv.}) is valid to all orders of $\epsilon$.
In the next section, we verify the propositions numerically using a three-level system (see Fig. \ref{fig:proof_3level}).

Now we replace $\mathcal{P}_{\rm inv}^{(\epsilon)}(t)$ by $\mathcal{P}_{\rm inv}^{(\epsilon)}$ and $\mathcal{J}_{\rm inv}^{(\epsilon)}(t)$ by zero in Eqs. (\ref{eq:TCL_full.x}).
As a result, we obtain
\begin{equation}
\begin{gathered}
    \dket{\rho(t)} = \mathcal{K}_{\rm TCL}^{(\epsilon)} \vec{x} (t) + e_{\mathcal{K}}(t), \\
    \frac{d}{dt} \vec{x}(t) = \mathcal{F}_{\rm TCL}^{(\epsilon)} \vec{x} (t) + e_{\mathcal{F}}(t),
\end{gathered}
\label{eq:TCL_full.x.error}
\end{equation}
where we have introduced 
\begin{equation}
    \mathcal{K}_{\rm TCL}^{(\epsilon)} = \hat{\mathcal{P}}_{\rm inv}^{(\epsilon)} \chi_R \ \ \ {\rm and} \ \ \ 
    \mathcal{F}_{\rm TCL}^{(\epsilon)} = \chi_L^\dagger \hat{\mathcal{L}} \hat{\mathcal{P}}_{\rm inv}^{(\epsilon)} \chi_R.
\label{eq:TCL_K.and.F}
\end{equation}
In Eqs. (\ref{eq:TCL_full.x.error}), $e_{\mathcal{K}}(t)$ and $e_{\mathcal{F}}(t)$ represent error terms due to the replacements.
From the exponential decaying factors in Eqs. (\ref{eq:TCL_P.relax}) and (\ref{eq:TCL_J.relax}), those are negligibly small in the long-time regime $t \gg \Delta^{-1}$.
Without the error terms, Eqs. (\ref{eq:TCL_full.x.error}) take similar forms to the equations defining $\mathcal{K}^{(\epsilon)}$
 and $\mathcal{F}^{(\epsilon)}$, namely, $\dket{\rho(t)} = \mathcal{K}^{(\epsilon)} \vec{x}(t)$ and $(d/dt) \vec{x}(t) = \mathcal{F}^{(\epsilon)} \vec{x}(t)$. 
This similarity implies that $\mathcal{K}_{\rm TCL}^{(\epsilon)}$ and $\mathcal{F}_{\rm TCL}^{(\epsilon)}$ correspond to $\mathcal{K}^{(\epsilon)}$ and $\mathcal{F}^{(\epsilon)}$, respectively.
When $\epsilon = 0$, we find $\mathcal{K}_{\rm TCL}^{(\epsilon = 0)} = \chi_R$, which yields $\mathcal{K}_{\rm TCL}^{(\epsilon = 0)} \vec{x} = \sum_s x_s \dket{r_{s}}$.
This confirms that $\mathcal{K}_{\rm TCL}^{(\epsilon = 0)}$ adheres to the boundary condition (\ref{eq:AE_K0}).
As outlined at the beginning, hence, the remaining task is to show the invariance condition (\ref{eq:AE_inv.cond.}), that is,
\begin{equation}
    \mathcal{K}_{\rm TCL}^{(\epsilon)} \mathcal{F}_{\rm TCL}^{(\epsilon)} = \hat{\mathcal{L}} \mathcal{K}_{\rm TCL}^{(\epsilon)}.
    \label{eq:TCL_inv.cond.}
\end{equation}
Inserting the definitions of $\mathcal{K}_{\rm TCL}^{(\epsilon)}$ and $\mathcal{F}_{\rm TCL}^{(\epsilon)}$, we obtain
\begin{gather*}
    \hat{\mathcal{L}} \mathcal{K}_{\rm TCL}^{(\epsilon)} - \mathcal{K}_{\rm TCL}^{(\epsilon)} \mathcal{F}_{\rm TCL}^{(\epsilon)}
    = \hat{\mathcal{L}} \hat{\mathcal{P}}_{\rm inv}^{(\epsilon)} \chi_R - \hat{\mathcal{P}}_{\rm inv}^{(\epsilon)} \hat{\mathcal{L}} \hat{\mathcal{P}}_{\rm inv}^{(\epsilon)} \chi_R \\
    = (\hat{\mathcal{I}} - \hat{\mathcal{P}}_{\rm inv}^{(\epsilon)}) \hat{\mathcal{L}} \hat{\mathcal{P}}_{\rm inv}^{(\epsilon)} \chi_R
    = 0,
\end{gather*}
where we have used $\mathcal{P}_{\rm inv}^{(\epsilon)} \mathcal{P}_{\rm inv} = \mathcal{P}_{\rm inv}^{(\epsilon)}$ in the first equality and Eq. (\ref{eq:TCL_proj.inv.}) in the last equality. 
Consequently, we confirm the invariance condition (\ref{eq:TCL_inv.cond.}), which ensures that $\mathcal{K}_{\rm TCL}^{(\epsilon)}$ and $\mathcal{F}_{\rm TCL}^{(\epsilon)}$ in the TCL master equation formulation agree with $\mathcal{K}^{(\epsilon)}$ and $\mathcal{F}^{(\epsilon)}$ in the geometric formulation.

\subsubsection*{Initial states in the invariant subspace  $\mathscr{M}^{(\epsilon)}$}

Next we examine scenarios with $\rho(0) \in \mathscr{M}^{(\epsilon)}$; the initial state $\rho(0)$ is in the invariant subspace with $\epsilon > 0$.
These scenarios are addressed in the geometric formulation, where the relations $\dket{\rho(t)} = \mathcal{K}^{(\epsilon)} \vec{x}(t)$ and $(d/dt) \vec{x}(t) = \mathcal{F}^{(\epsilon)} \vec{x}(t)$ hold for all times. 
We here confirm these relations in the TCL master equation formulation.
To this end, we adopt $\mathcal{P} = \mathcal{P}_{\rm inv}^{(\epsilon)}$.
From Proposition \ref{prop:proj}, $\rho(0)$ in the present discussion satisfies $\rho(0) = \mathcal{P}_{\rm inv}^{(\epsilon)} \rho(0)$ or, equivalently, $\mathcal{Q}_{\rm inv}^{(\epsilon)} \rho(0) = 0$ with $\mathcal{Q}_{\rm inv}^{(\epsilon)} = \mathcal{I} - \mathcal{P}_{\rm inv}^{(\epsilon)}$.
Furthermore, the relation (\ref{eq:TCL_proj.inv.}), which reads $\mathcal{Q}_{\rm inv}^{(\epsilon)} \mathcal{L} \mathcal{P}_{\rm inv}^{(\epsilon)} = 0$, ensures that the corresponding $\Sigma(t)$ defined in Eq. (\ref{eq:TCL_Sigma_general}) vanishes.
We hence consider Eqs. (\ref{eq:TCL_Q.rho.2nd}) and (\ref{eq:TCL_TCL.2nd}), where the terms involving $\mathcal{Q}_{\rm inv}^{(\epsilon)} \rho(0)$ vanish from the above discussion.
Consequently, Eqs. (\ref{eq:TCL_Q.rho.2nd}) and (\ref{eq:TCL_TCL.2nd}) in the vector representation simplify to $\dket{\rho(t)} = \hat{\mathcal{P}}_{\rm inv}^{(\epsilon)} \dket{\rho(t)}$ and
\begin{equation*}
    \frac{d}{dt} \hat{\mathcal{P}}_{\rm inv}^{(\epsilon)} \dket{\rho (t)} = \hat{\mathcal{P}}_{\rm inv}^{(\epsilon)} \hat{\mathcal{L}} \hat{\mathcal{P}}_{\rm inv}^{(\epsilon)} \dket{\rho (t)},
\end{equation*}
respectively.
Applying $\mathcal{P}_{\rm inv}$ from the left side to the latter equation and using $\mathcal{P}_{\rm inv} \mathcal{P}_{\rm inv}^{(\epsilon)} = \mathcal{P}_{\rm inv}$, we obtain 
\begin{equation}
    \frac{d}{dt} \hat{\mathcal{P}}_{\rm inv} \dket{\rho(t)}  = \hat{\mathcal{P}}_{\rm inv} \hat{\mathcal{L}} \hat{\mathcal{P}}_{\rm inv}^{(\epsilon)} \dket{\rho(t)}.
    \label{eq:TCL_reduced.dynamics}
\end{equation}
Expressing with the parameter $\vec{x}(t)$ and inserting the definitions of $\mathcal{K}_{\rm TCL}^{(\epsilon)}$ and $\mathcal{F}_{\rm TCL}^{(\epsilon)}$, we obtain 
\begin{equation}
    \dket{\rho(t)} = \mathcal{K}_{\rm TCL}^{(\epsilon)} \vec{x} (t), \ \
    \frac{d}{dt} \vec{x}(t) = \mathcal{F}_{\rm TCL}^{(\epsilon)} \vec{x}(t).
\label{eq:TCL_full.x.}
\end{equation}
In contrast to Eqs. (\ref{eq:TCL_full.x.error}), Eqs. (\ref{eq:TCL_full.x.}) hold true for all times $t$ without error terms.
This is due to the absence of the fast relaxation phase when initially placed in $\mathscr{M}^{(\epsilon)}$.
This analysis confirms that the TCL master equation formulation offers a consistent description with the geometric formulation.

\subsubsection*{Remarks}

Five remarks are in order.
\begin{enumerate}[label=(\alph*)]

    \item Proposition \ref{prop:relax} characterizes the relaxation behavior of $\mathcal{P}_{\rm inv}^{(\epsilon)} (t) - \mathcal{P}_{\rm inv}^{(\epsilon)}$ and $\mathcal{J}_{\rm inv}^{(\epsilon)} (t)$ in the long-time regime.
    The former has been widely recognized in numerous practical applications.
    In contrast, studies on the latter remain scarce. This is primarily because $\mathcal{Q} \rho(0) = 0$ is commonly assumed in the literature.
    In the geometric picture, states invariably converge to $\mathscr{M}^{(\epsilon)}$, irrespective of the initial conditions. From this, it is expected that the terms that explicitly depend on the initial condition should  disappear in the long-time regime. The exponential decaying factor in Eq. (\ref{eq:TCL_J.relax}) provides a theoretical basis for this expectation.
    Proposition \ref{prop:proj}, on the other hand, offers a geometric interpretation of $\mathcal{P}_{\rm inv}^{(\epsilon)}$, which has not been recognized in the literature to our knowledge. 
    Equation (\ref{eq:TCL_proj.inv.}) can be seen as an extension of the relation $\mathcal{Q}_{\rm inv} \mathcal{L}_0 \mathcal{P}_{\rm inv} = 0$ to nonzero $\epsilon$.

    \item The geometric formulation has mostly been applied to the cases with $\lambda_s = 0$.
    This limitation is partly because of the complexities involved in solving the invariance condition when $\lambda_s$ is nonzero, as detailed in Appendix \ref{app:inv.cond.}.
    The extension to nonzero $\lambda_s$ has recently been achieved in Ref. \cite{Angela24}, where the invariance condition was solved by regarding it as a Sylvester equation for superoperators.
    On the other hand, the procedure of computing the maps $\mathcal{K}_{\rm TCL}^{(\epsilon)}$ and $\mathcal{F}_{\rm TCL}^{(\epsilon)}$ remains consistent irrespective of the values of $\lambda_s$.

    \item The geometric formulation provides a description of the dynamics within $\mathscr{M}^{(\epsilon)}$ but fails to account for the fast relaxation phase.
    In contrast, the TCL master equation formulation enables a straightforward description of the fast relaxation phase by simply considering a finite time $t$ in Eq. (\ref{eq:TCL_full.rho}).
    This highlights the broader applicability of the TCL master equation formulation.
    Moreover, neglecting the short-time dynamics can result in pathological behavior in the long-time regime, and this can be addressed using the TCL master equation. This issue is further elucidated through an example in Sec. \ref{Demo_bipartite}.

    \item To evaluate the maps $\mathcal{F}_{\rm TCL}^{(\epsilon)}$ and $\mathcal{K}_{\rm TCL}^{(\epsilon)}$ using perturbative methods, it is necessary to first compute a perturbation expansion of $\mathcal{P}_{\rm inv}^{(\epsilon)}$.
    In Appendix \ref{app:perturb}, we provide explicit formulas up to the third-order of $\epsilon$.
    By identifying the specific operation of $\mathcal{P}_{\rm inv}$ for the system under investigation, the map $\mathcal{P}_{\rm inv}^{(\epsilon)}$ can be straightforwardly evaluated using the provided formulas.
    In the next section, we demonstrate this procedure with an example.
    
    \item For bipartite systems that exhibit a unique steady state in a subsystem (details are elaborated in Sec. \ref{Demo_bipartite}), a recent study introduced an alternative formulation of adiabatic elimination based on the projection method \cite{Saideh20}.
    The authors employed the superoperator $\mathcal{P}$ that projects onto a product state including the unique steady state of the subsystem.
    This definition of $\mathcal{P}$ agrees with the one derived from Eq. (\ref{eq:TCL_P.def}) as we demonstrate in Sec. \ref{Demo_bipartite}.
    Instead of the TCL master equation, the authors focused on the Nakajima-Zwanzig equation (\ref{eq:TCL_NZ}).
    By applying the Laplace transform, they derived an expression of the generator for $\mathcal{P} \rho(t)$ in a long-time domain (see Appendix \ref{app:Laplace} for more details).
    Although our starting point aligns with the one in Ref. \cite{Saideh20}, we identify three key distinctions.
    Firstly, as in most adiabatic elimination formulations, there is no consideration of a map analogous to $\mathcal{K}^{(\epsilon)}$.
    This limits their formulation to only providing the reduced dynamics $\mathcal{P} \rho(t)$, without the capability to extract the total density operator $\rho(t)$.
    Secondly, it was assumed that the initial condition $\rho(0)$ satisfies $\mathcal{Q} \rho(0) = 0$.
    In contrast, our formulation does not require such an assumption as the term involving $\rho(0)$ becomes small anyway in the long-time regime as substantiated by Eq. (\ref{eq:TCL_J.relax}).
    Finally, as we show in Appendix \ref{app:Laplace}, the generator for $\mathcal{P} \rho(t)$ introduced in Ref. \cite{Saideh20} generally does not agree with that of the geometric formulation.
    On the other hand, the generator in our formulation is consistent with the geometric formulation.

\end{enumerate}

\section{Demonstrations}
\label{Demo}

In this section, we apply the TCL master equation framework of adiabatic elimination to two specific examples.
In Sec. \ref{Demo_prop}, we examine a three-level system to numerically verify the propositions outlined in the previous section.
In Sec. \ref{Demo_bipartite}, we explore a bipartite system to illustrate the practical procedure of adiabatic elimination, providing a concrete example of the Rabi model with a strongly damped oscillator mode.

\subsection{Numerical verification of the propositions}
\label{Demo_prop}

\begin{figure*}[t]
  \includegraphics[keepaspectratio, scale=0.6]{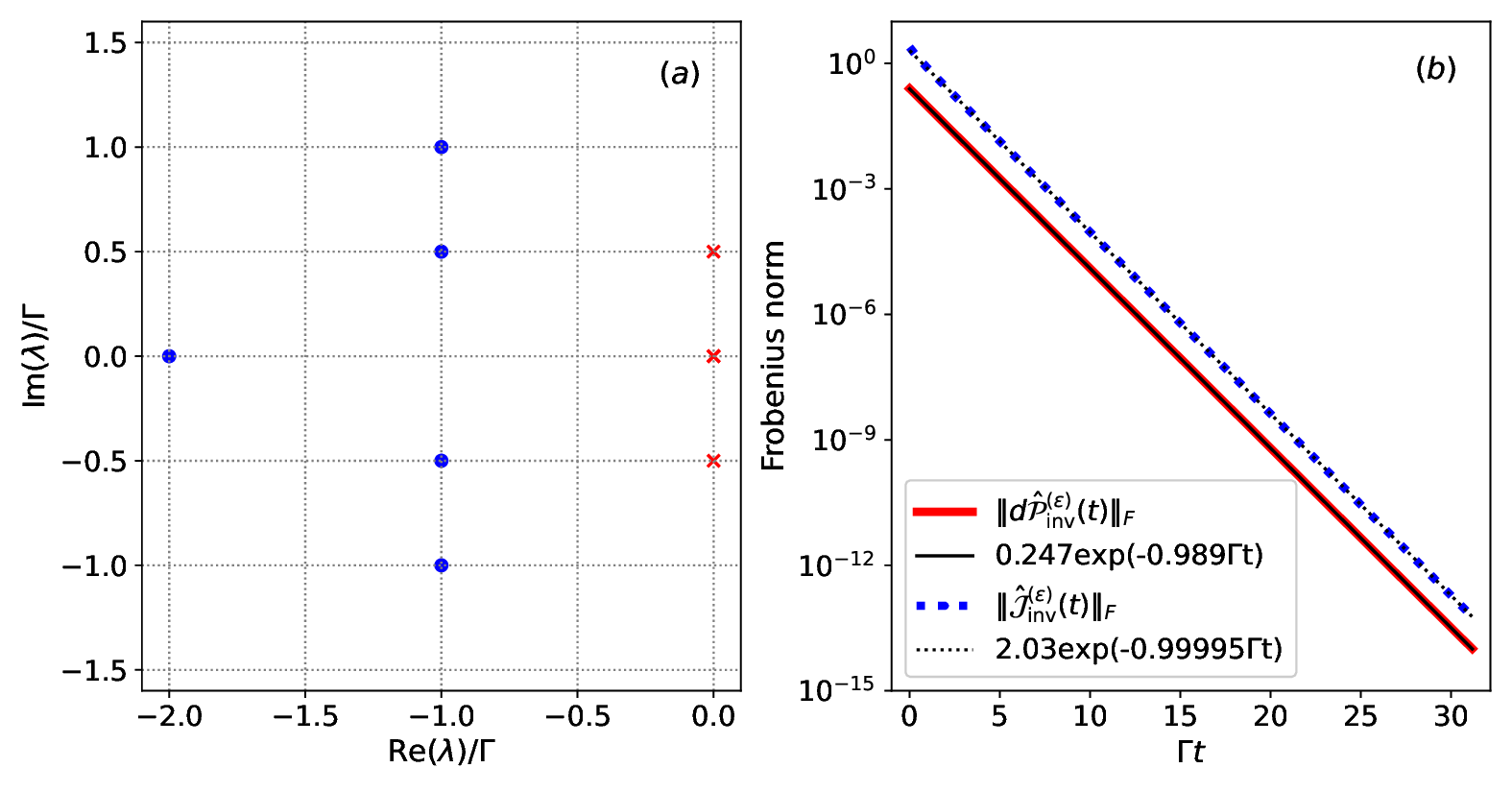}
  \caption{
  (a) Eigenvalues of $\mathcal{L}_0$ for the three-level system defined by Eq. (\ref{eq:proof_L0.3level}).
  Consistent with Fig. \ref{fig:AE_gap}, the red crosses denote the surviving modes, whereas the blue circles represent the fast relaxation modes.
  (b) Exponential decay of the Frobenius norms, $\| d \hat{\mathcal{P}}_{\rm inv}^{(\epsilon)} (t) \|_F$ (thick solid red line) and $\| \hat{\mathcal{J}}_{\rm inv}^{(\epsilon)} (t) \|_F$ (thick dotted blue line), for the three-level system.
  The thin solid black line and thin dotted black line represent the fitting results for $\| d \hat{\mathcal{P}}_{\rm inv}^{(\epsilon)} (t) \|_F$ and $\| \hat{\mathcal{J}}_{\rm inv}^{(\epsilon)} (t) \|_F$, respectively. 
  }
  \label{fig:proof_3level}
\end{figure*}

From Propositions \ref{prop:relax} and \ref{prop:proj}, we have found that, for small $\epsilon$, the magnitudes of $\hat{\mathcal{P}}_{\rm inv}^{(\epsilon)} (t) - \sum_{s s'} \dket{r_{s}^{(\epsilon)}} [N^{-1}]_{s s'} \dbra{l_{s'}} \equiv d \hat{\mathcal{P}}_{\rm inv}^{(\epsilon)} (t)$ and $\hat{\mathcal{J}}_{\rm inv}^{(\epsilon)}$ decay exponentially with time. 
Here we verify such behavior numerically. 
To this end, we consider a three-level system with spontaneous decay as an example. 
Introducing the orthonormal tree-level states $\{ \ket{0}, \ket{1}, \ket{e} \}$, we consider a GKSL generator $\mathcal{L} = \mathcal{L}_0 + \epsilon \mathcal{L}_1$ with \cite{Reiter12}
\begin{equation}
    \begin{gathered}
        \mathcal{L}_0 \rho = - i [ \, \omega_1 \ketbra{1}{1} + \omega_e \ketbra{e}{e}, \, \rho] \\
        + 2 \Gamma_0 \mathcal{D}[\ketbra{0}{e}] \rho + 2 \Gamma_1 \mathcal{D}[\ketbra{1}{e}] \, \rho,
    \end{gathered}
    \label{eq:proof_L0.3level}
\end{equation}
and 
\begin{equation*}
    \epsilon \mathcal{L}_1 \rho = - i \left[ g_0 \ketbra{0}{e} + g_1 \ketbra{1}{e} + (H.c.), \, \rho \right],
\end{equation*}
where $\omega_1$ ($\omega_e$) is the frequency associated with the excitation energy of the state $\ket{1}$ ($\ket{e}$), $\Gamma_{0}$ ($\Gamma_1$) is the rate at which the spontaneous decay from the excited state $\ket{e}$ to the state $\ket{0}$ ($\ket{1}$) occurs, $g_0$ ($g_1$) is the strength of the coupling between the states $\ket{e}$ and $\ket{0}$ ($\ket{1}$), and $H.c.$ denotes the Hermitian conjugate of the preceding terms.
The dissipator superoperator is introduced as $\mathcal{D}[L] \rho = L \rho L^\dagger - (L^\dagger L \rho + \rho L^\dagger L)/2$.
The free part $\mathcal{L}_0$ can be diagonalized analytically. With $\Gamma = \Gamma_0 + \Gamma_1$, we arbitrarily set $\omega_e = \Gamma$, $\omega_1 = 0.5 \Gamma$. In this case, the spectrum of $\mathcal{L}_0$ is given by Fig. \ref{fig:proof_3level} (a).
We see that the gap is given by $\Delta = \Gamma$.

The quantities $\mathcal{P}_{\rm inv}^{(\epsilon)}(t)$ and $\mathcal{J}_{\rm inv}^{(\epsilon)}(t)$ can be computed numerically from their definitions. In practice, it is preferable to utilize the equivalent expressions given in Eqs. (\ref{eq:proof_P.inv.eps}) and (\ref{eq:proof_J.inv.eps}), as they circumvent the need for time integration. The projection $\sum_{s s'} \dket{r_{s}^{(\epsilon)}} [N^{-1}]{s s'} \dbra{l_{s'}}$ can be evaluated through a numerical diagonalization of the generator $\hat{\mathcal{L}}$.

Employing the Frobenius norm $\| \|_F$, Fig. \ref{fig:proof_3level} (b) shows the results with $\Gamma_0 = \Gamma_1 = 0.5 \Gamma$ and $g_0 = g_1 = 0.1 \Gamma$, in which case we have $\epsilon/\Delta \simeq 0.1$.
For the matrix representation of superoperators, we employed a vectorization representation presented in Appendix C of Ref. \cite{TESR23}.
The exponential decay behavior can be observed clearly.
The thin lines in Fig. \ref{fig:proof_3level} (b) show the fitting results.
We find that the slopes in the logarithmic plot are close to unity and that the ratio $\| d \hat{\mathcal{P}}_{\rm inv}^{(\epsilon)} (t) \|_F / \| \hat{\mathcal{J}}_{\rm inv}^{(\epsilon)} (t) \|_F \simeq 0.247 / 2.03 \simeq 0.122$ is in the order of $\epsilon/\Delta$. 
These are in agreement with the order estimates in Proposition \ref{prop:relax}.

\subsection{Bipartite systems}
\label{Demo_bipartite}

For more complex systems, where direct numerical evaluation becomes infeasible, perturbation calculations are needed.
In the TCL master equation formulation, the basic procedure consists in three steps: 
\begin{enumerate}
    \item Identifying $\mathcal{L}_0$ and $\mathcal{L}_1$ in the dynamics.
    \item Identifying the surviving modes by solving the zero order problem.
    \item Specifying the operation of the projection $\mathcal{P}_{\rm inv}$ defined by Eq. (\ref{eq:TCL_P.def}).
\end{enumerate}
To demonstrate these steps, we here consider adiabatic elimination for a bipartite system, which serves as a prototypical model for reservoir engineering. 
We show that the projection $\mathcal{P}_{\rm inv}$ is given by the partial trace over the eliminated subsystem [see Eq. (\ref{eq:projection_operation})].
Furthermore, we apply this scheme to the Rabi model and discuss the positivity of the reduced dynamics.

Consider a bipartite system consisting of two subsystems $A$ and $B$, the dimensions of which are denoted as $d_A$ and $d_B$, respectively.
Suppose that the density operator $\rho$ obeys a master equation of the form
\begin{equation}\label{eq:Demo_bipartite.L}
    \frac{d}{dt} \rho (t) = \mathcal{L}\rho(t),\quad \mathcal{L} = \mathcal{L}_A \otimes \mathcal{I}_B + \mathcal{I}_A \otimes \mathcal{L}_B + \mathcal{L}_{\rm{int}},
\end{equation}
where, for $\xi = A, B$, $\mathcal{I}_\xi$ is the identity superoperator on $\xi$, $\mathcal{L}_\xi$ describes the internal dynamics on $\xi$, and $\mathcal{L}_{\rm{int}}$ represents the interaction between $A$ and $B$.
In what follows, we assume that $\mathcal{L}_A$ has a unique steady state and $\mathcal{L}_B$ generates unitary dynamics as $\mathcal{L}_B \rho  = - i [H_B, \rho]$.

The first step is to identify $\mathcal{L}_0$ and $\epsilon \mathcal{L}_1$ in the dynamics. This procedure depends on the specific physical setting to be investigated.
We here consider cases where a strongly dissipative subsystem $A$ is weakly coupled to another subsystem $B$ via $\mathcal{L}_{\rm int}$.
Owing to the weak coupling assumption, we can treat $\mathcal{L}_{\rm int}$ as perturbation.
Therefore, we set $\mathcal{L}_0 = \mathcal{L}_A \otimes \mathcal{I}_B + \mathcal{I}_A \otimes \mathcal{L}_B$ and $\epsilon \mathcal{L}_1 = \mathcal{L}_{\rm{int}}$.

The second step is to identify the surviving modes by solving the eigenvalue problem of $\mathcal{L}_0$ for them.
Suppose the eigenvalue problem of $\mathcal{L}_A$ is solved as
\begin{equation}
     (\hat{\mathcal{L}}_A - \lambda_{A,i} \hat{\mathcal{I}}_A ) \dket{r_{A,i}} = 0, \ \ \dbra{l_{A,i}} (\hat{\mathcal{L}}_A - \lambda_{A,i}  \hat{\mathcal{I}}_A) = 0,
\label{eq:eigenproblem_example}
\end{equation}
for $i = 1,2,\dots,d_A^2$.
The assumption of a unique steady state, denoted by $\bar{\rho}_A$, then indicates $\lambda_{A,1} = 0$, $r_{A,1} = \bar{\rho}_A$, $l_{A,1} = I_A$ ($I_\xi$ the identity operator on $\xi$), and ${\rm Re} \lambda_{A,i>1} < 0$.
The form of the left eigenvector $l_{A,1}$ follows from $\dbra{I_A} \hat{\mathcal{L}}_A \rightarrow {\rm tr}_A \mathcal{L}_A = 0$, with ${\rm tr}_A$ the trace operation over $A$.
Note that the decay rate in $A$ is characterized by $\min_{i > 1} |{\rm Re} \lambda_{A,i}|$.
For the subsystem $B$, suppose that the eigenvalue problem of $H_B$ is solved as
\begin{equation*}
    (H_B - \Omega_{B,m} I_B) \ket{b_m} = 0,
\end{equation*}
for $m = 1,2,\dots,d_B$. The (right and left) eigenvectors of $\mathcal{L}_B$ then read $\ketbra{b_m}{b_n} \equiv E_{B,mn}$ with the eigenvalue $i (\Omega_{B,n} - \Omega_{B,m}) \equiv i \Omega_{B,mn}$. 
Assuming $\{ \ket{b_m} \}_{1 \leq m \leq d_B}$ to be an orthonormal basis on $B$, we find the orthonormal relation $\dbraket{E_{B,mn}}{E_{B,pq}} = \delta_{m,p}\delta_{n,q}$ and the resolution of identity $\hat{\mathcal{I}}_B = \sum_{m,n=1}^{d_B} \dketbra{E_{B,mn}}{E_{B,mn}}$.

Given these, the eigenvalue problem of $\hat{\mathcal{L}}_0$ can now be solved formally.
The right and left eigenvectors are given by $\{ \dket{r_{A,i}} \otimes  \dket{E_{B,mn}} \}^{1 \leq i \leq d_A^2}_{1 \leq m, n \leq d_B}$ and $\{ \dbra{l_{A,i}} \otimes  \dbra{E_{B,mn}} \}^{1 \leq i \leq d_A^2}_{1 \leq m, n \leq d_B}$, respectively, and the eigenvalues read $\lambda_{i,m,n} = \lambda_{A,i} + i \Omega_{B,mn}$.
Note that $\min_{i > 1,m,n} |{\rm Re} \lambda_{i,m,n}| = \min_{i > 1} |{\rm Re} \lambda_{A,i}|$, where $\min_{i > 1} |{\rm Re} \lambda_{A,i}|$ is the decay rate in $A$ as mentioned above.
The strong dissipation assumption on $A$ implies that this decay rate is much faster compared to the typical scale of $\epsilon \mathcal{L}_1$.
In this case, we can take the modes $(i=1,m,n)_{1 \leq m, n \leq d_B}$ as the surviving modes.
The set of eigenvalues bears a resemblance to Fig. \ref{fig:AE_gap}, with the red crosses on the imaginary axis corresponding to $\{ \lambda_{i=1,m,n} \}_{1 \leq m, n \leq d_B}$ and with the gap given by $\Delta = \min_{i > 1} |{\rm Re} \lambda_{A,i}|$.

The last step is to specify the operation of the projection  $\mathcal{P}_{\rm inv}$.
The general definition of $\mathcal{P}_{\rm inv}$ is given by Eq. (\ref{eq:TCL_P.def}). 
Since the right and left eigenvectors associated with the surviving modes are $\{ \dket{r_{A,1}} \otimes  \dket{E_{B,mn}} \}_{1 \leq m, n \leq d_B}$ and $\{ \dbra{l_{A,1}} \otimes  \dbra{E_{B,mn}} \}_{1 \leq m, n \leq d_B}$, respectively, we obtain
\begin{gather}
    \hat{\mathcal{P}}_{\rm inv} = \sum_{m,n = 1}^{d_B} \dket{r_{A,1}} \otimes \dket{E_{B,mn}} \dbra{l_{A,1}} \otimes  \dbra{E_{B,mn}} \nonumber \\
    = \dketbra{\bar{\rho}_A}{I_A} \otimes \hat{\mathcal{I}}_B,
    \label{eq:projection_vectorized}
\end{gather}
the operation of which reads
\begin{equation}
    \mathcal{P}_{\rm inv} \rho = \bar{\rho}_A \otimes {\rm tr}_A (\rho).
    \label{eq:projection_operation}
\end{equation}
We note that the abstract definition (\ref{eq:TCL_P.def}) naturally yields Eq. (\ref{eq:projection_operation}), which is commonly assumed in the study of bipartite systems. 

Two remarks are in order.
\begin{enumerate}

\item[$(\alpha)$] As noted at the end of Sec. \ref{AE}, the invariance condition (\ref{eq:AE_inv.cond.}) has multiple solutions due to the nonuniqueness of parametrizing the invariant subspace $\mathscr{M}^{(\epsilon)}$.
In this article, we impose the condition (\ref{eq:AE_gauge}), which is equivalent to the parameter choice given by $x_s = \dbraket{l_s}{\rho}$.
In the current setting, the left eigenvectors associated with the surviving modes are $\{ \dbra{l_{A,1}} \otimes  \dbra{E_{B,mn}} \}_{1 \leq m, n \leq d_B}$.
Accordingly, the parameters read
\begin{equation*}
    x_{mn}= \dbraket{ E_{B,mn} }{ \rho_B },
\end{equation*}
with $\rho_B = {\rm tr}_A \rho$ the reduced density operator on $B$. 
Using the resolution of identity, we find
\begin{equation}
    \rho_B = \sum_{m,n=1}^{d_B} x_{mn}  E_{B,mn}.
    \label{eq:Demo_rhoB}
\end{equation}
Therefore, in bipartite systems, the gauge condition (\ref{eq:AE_gauge}) naturally yields the parametrization via the reduced density operator.

\item[$(\beta)$] When the internal dynamics of $B$ have a much slower timescale compared to $\Delta$, $\mathcal{L}_B$ can be treated as perturbation. 
We can then split $\mathcal{L}$ as $\mathcal{L}_0 = \mathcal{L}_A \otimes \mathcal{I}_B$ and $\epsilon \mathcal{L}_1 = \mathcal{I}_A \otimes \mathcal{L}_B 
+ \mathcal{L}_{\rm{int}}$.
In this case, the right and left eigenvectors are given as above with $\{ E_{B,mn} \}_{1 \leq m,n \leq d_B}$ being arbitrary orthonormal operator basis. 
Hence, the operation of $\mathcal{P}_{\rm inv}$ is similarly given by Eq. (\ref{eq:projection_operation}).
For systems dictated by a GKSL equation, such cases were investigated in Ref. \cite{Azouit17}.
In particular, the authors proved that the second-order reduced dynamics of $\rho_B$ are always given in the GKSL form.
When $\mathcal{L}_B$ has a comparable timescale to $\Delta$ and needs to be incorporated in $\mathcal{L}_0$, however, the GKSL form is no longer guaranteed as we see below.

\end{enumerate}

\subsubsection*{Rabi model}

Once the operation of $\mathcal{P}_{\rm inv}$ is specified, the evaluation of $\mathcal{F}^{(\epsilon)}_{\rm TCL}$ and $\mathcal{K}^{(\epsilon)}_{\rm TCL}$ is straightforward.
We here present explicit computations for the Rabi model including a damped oscillator mode.
The dissipative subsystem $A$ is the system of an oscillator mode (hereafter referred to as photon) with dynamics dictated by a GKSL equation
\begin{equation}
    \mathcal{L}_A \rho  = - i \omega_{\rm ph} [a^\dagger a, \rho] +  \kappa \mathcal{D}[a] \rho,
\label{eq:Demo_Rabi.LA}
\end{equation}
where $\omega_{\rm ph}$ is the frequency of the photon and $\kappa$ is the single photon loss rate. 
The subsystem $B$ is a qubit system spanned by the orthonormal basis $\{ \ket{g}, \ket{e} \}$.
The internal Hamiltonian of $B$ is assumed to be 
\begin{equation*}
    H_B = \frac{\omega_{\rm eg}}{2} \sigma_z,
\end{equation*}
with $\omega_{\rm eg}$ the energy difference between the two levels and $\sigma_z = \ketbra{e}{e} - \ketbra{g}{g}$. 
Hence, $\mathcal{L}_B$ reads 
\begin{equation}
    \mathcal{L}_B \rho = -  \frac{i \omega_{\rm eg}}{2} \left[ \sigma_z, \rho \right].
\label{eq:Demo_Rabi.LB}
\end{equation}
The two subsystems are coupled by the Rabi interaction
\begin{equation}
    \mathcal{L}_{\rm int} \rho = - i g \left[ \left(a^\dagger + a \right) \otimes \sigma_x , \rho \right],
\label{eq:Demo_Rabi.Lint}
\end{equation}
with $\sigma_x = \sigma_+ + \sigma_-$ and $\sigma_+ = \sigma_-^\dagger = \ketbra{e}{g}$.

The Rabi model is an infinite-dimensional system to which some results in Appendix \ref{app:proof} do not directly apply.
While a rigorous extension to infinite-dimensional systems is beyond the scope of this work, such systems are crucial for quantum technology applications.
To examine applicability, we numerically analyze the effect of photon space truncation for the Rabi model in Appendix \ref{app:infdim}.
The results suggest that the equivalence between the geometric and TCL master equation formulations remains valid in this case.
Accordingly, we assume the equivalence in the following discussions and adopt the TCL master equation formulation.

With $\mathcal{L}_A$ given by Eq. (\ref{eq:Demo_Rabi.LA}), we find that $\Delta = \min_{i > 1} |{\rm Re} \lambda_{A,i}|  = \kappa / 2$.
Accordingly, the assumption of weak coupling reads $g / \kappa \ll 1$.
Under this condition, the maps $\mathcal{F}_{\rm TCL}^{(\epsilon)}$ and $\mathcal{K}_{\rm TCL}^{(\epsilon)}$ can be evaluated perturbatively.
Detailed calculations are provided in Appendix \ref{app:Rabi}.
In what follows, we present the results and focus on their physical significance.

For the parametrization $\rho_B$ in Eq. (\ref{eq:Demo_rhoB}), the reduced dynamics up to the second-order expansion read $(d/dt) \rho_B (t) = \mathcal{F}_{\rm TCL}^{(\epsilon)} \rho_B (t)$ with [see Eq. (\ref{eq:Rabi_FTCL})]
\begin{equation*}
\begin{gathered}
    \mathcal{F}_{\rm TCL}^{(\epsilon)} \rho_B = - \frac{i}{2} \Big\{ \omega_{\rm eg} + g^2 {\rm Im}(1 / \gamma_+ - 1 / \gamma_-) \Big\} [\sigma_z, \rho_B] \\ 
    + g^2 \sum_{j,k = \pm} K_{jk} \Bigg[\sigma_j \rho_B \sigma_k^\dagger -\frac{\sigma_k^\dagger \sigma_j \rho_B + \rho_B \sigma_k^\dagger \sigma_j}{2} \Bigg],
\end{gathered}
\end{equation*}
with $\gamma_{\pm} = (\kappa/2) + i (\omega_{\rm ph} \pm \omega_{eg})$ and $K_{jk} = 1/\gamma_j + 1/\gamma_k^*$. 
This result agrees with the geometric approach based on a Sylvester equation \cite{Angela24} and with the Redfield equation derived in Ref. \cite{Damanet19}.

The eigenvalues of the coefficient matrix in front of the dissipator, $K$, read
\begin{equation}
    \frac{{\rm tr}(K)}{2} \left[ 1 \pm \sqrt{1 + \Bigg( \frac{4 \omega_{\rm eg}}{|\gamma_+ \gamma_-| \ {\rm tr}(K)} \Bigg)^2 } \ \right],
    \label{eq:Demo_K.eigval}
\end{equation}
with ${\rm tr}(K) = \kappa / |\gamma_+|^2 + \kappa / |\gamma_-|^2 > 0$.
If $\omega_{\rm eg} \ne 0$, one of the eigenvalues becomes negative.
This implies that the second-order generator is in a non-GKSL form violating complete positivity of the evolution \cite{GKS,Lindblad}.
This issue was pointed out in Ref. \cite{Damanet19}, where the authors showed that additional approximations on the generator to obtain the GKSL form lead to qualitatively incorrect dynamics.
An approach that ensures complete positivity was developed in Ref. \cite{Jager22}, where a master equation in the GKSL form was derived based on the Schrieffer-Wolff transformation.

Here we propose an alternative way to ensure complete positivity of the reduced dynamics. In Refs. \cite{Maniscalco04,Whitney08,Hartmann20}, the authors demonstrated, for composite Hamiltonian systems, that taking into account the time dependence of coefficients maintains the positivity of the density operator even in the short-time regime.
Following these insights, we propose to use
\begin{equation}
    \mathcal{F}_{\rm TCL}^{(\epsilon)} (t) \equiv \chi_L^\dagger \hat{\mathcal{L}} \hat{\mathcal{P}}_{\rm inv}^{(\epsilon)} (t) \chi_R
    \label{eq:Demo_FTCL.td},
\end{equation}
as the generator, rather than the asymptotic one $\mathcal{F}_{\rm TCL}^{(\epsilon)} = \lim_{t \to \infty} \mathcal{F}_{\rm TCL}^{(\epsilon)} (t)$, when the positivity violation becomes an issue.
Assuming $\mathcal{Q}_{\rm inv} \rho(0) = 0$ as in previous studies, the modified reduced dynamics are now given by $(d/dt) \rho_B (t) = \mathcal{F}_{\rm TCL}^{(\epsilon)} (t) \rho_B (t)$.
We numerically demonstrate in Appendix \ref{app:Rabi_CP} that this evolution is completely positive at all times (see Fig. \ref{fig:Rabi_Choi}), and we expect this to hold generally, as long as the coupling is weak enough so that the second-order approximation is justified.
We emphasize that the time-dependent generator cannot be derived within the geometric approach, while it can be straightforwardly evaluated in the TCL master equation formulation.

Once the evolution of the parameter $\rho_B(t)$ is determined, the density operator of the bipartite system can be obtained as $\rho(t) = \mathcal{K}_{\rm TCL}^{(\epsilon)} \rho_B(t)$ with,
up to the order of $\epsilon^2$, the map $\mathcal{K}_{\rm TCL}^{(\epsilon)}$ given by [see Eq. (\ref{eq:Rabi_KTCL})]
\begin{equation*}
\begin{gathered}
    \mathcal{K}_{\rm TCL}^{(\epsilon)} \rho_B = (I + W) (\ketbra{0}{0} \otimes \rho_B) (I + W)^\dagger \\
    - g^2 (I_A \otimes \sigma_\gamma) (\ketbra{0}{0} \otimes \rho_B) (I_A \otimes \sigma_\gamma)^\dagger.
\end{gathered}
\end{equation*}
with $I$ the identity operator on the total space, $\sigma_\gamma = \sigma_- / \gamma_- + \sigma_+ / \gamma_+$, and $W = - ig a^\dagger \otimes \sigma_\gamma - g^2 / (\kappa + 2 i \omega_{\rm ph}) (a^\dagger)^2 \otimes (\sigma_- \sigma_+ / \gamma_+ + \sigma_+ \sigma_- / \gamma_-)$.
Unlike the reduced dynamics, this result has not been derived in the literature to our knowledge.
Note that $\mathcal{K}_{\rm TCL}^{(\epsilon)}$ is not a Kraus map due to the negative sign in the second line.
This can be interpreted as a signature of quantum correlation built up in states in $\mathscr{M}^{(\epsilon)}$ \cite{TESR23}.

\section{Concluding remarks}
\label{summary}

In this study, we have established the TCL master equation approach to adiabatic elimination. 
The standard geometric approach characterizes the dynamics on the lower-dimensional invariant subspace $\mathscr{M}^{(\epsilon)}$ through two maps; $\mathcal{F}^{(\epsilon)}$, describing the time evolution on $\mathscr{M}^{(\epsilon)}$ and $\mathcal{K}^{(\epsilon)}$, connecting the degrees of freedom on $\mathscr{M}^{(\epsilon)}$ to the total density operator.
By expressing these maps within the TCL master equation framework, we have demonstrated that our approach yields results identical to those of the geometric approach, establishing the equivalence of the two formulations.
This equivalence is proven through two original Propositions.
Proposition \ref{prop:relax} guarantees the exponential suppression of time-dependent terms appearing in the two maps.
Proposition \ref{prop:proj} shows that the image of the projection $\mathcal{P}_{\rm inv}^{(\epsilon)}$ [defined below Eq. (\ref{eq:TCL_full.x})] is in fact the invariant subspace $\mathscr{M}^{(\epsilon)}$, providing geometric insights into the TCL master equation formalism.
These propositions are proved in Appendix \ref{app:proof} and are verified numerically in Sec. \ref{Demo_prop}.
In Sec. \ref{Demo_bipartite}, we demonstrate how our formulation of adiabatic elimination allows for seamless treatment of cases with rapid unitary dynamics in $\mathscr{M}^{(\epsilon = 0)}$
and of the transient regime with the two maps still being time-dependent.
This underscores the practical value of our formulation.

Overall, this study broadens the applicability of the adiabatic elimination method, by recasting it in the widely recognized TCL master equation framework.
Given that the TCL master equation has been extensively studied from various perspectives, this opens up numerous possibilities for applying already developed techniques to adiabatic elimination problems. For instance, diagrammatic techniques that simplify the evaluation of higher-order contributions, such as those developed for computing the generator of the dynamics \cite{Gasbarri18} and for steady states \cite{Milena21, Ferguson21}, might now be employed. Additionally, a difference in how the two maps are defined between the two approaches may enable new investigations. 
In contrast to the geometric approach, where the two maps are the solutions to the invariance condition (\ref{eq:AE_inv.cond.}), our approach offers concrete expressions of these maps, as shown in Appendix \ref{app:proof}.
These expressions could facilitate numerical studies of complex systems, where the dynamics generated by the free part $\mathcal{L}_0$ are not analytically accessible.

\begin{acknowledgments}
    We are grateful to Alain Sarlette, Pierre Rouchon, Mazyar Mirrahimi, Alexandru Petrescu, and Chikako Uchiyama for valuable discussions during the preparation of this article.
    M.T. was supported by JSPS KAKENHI Grant Number JP23KJ1157.
    A.R. was supported by ANR grants HAMROQS and MECAFLUX (French Research Agency), and by Plan France 2030 through the project ANR-22-PETQ-0006.
\end{acknowledgments}

\section*{Data Availability}
The data that support the findings of this article are openly available \cite{Data}.

\appendix

\section{Perturbation solution to the invariance condition}
\label{app:inv.cond.}

In this appendix, we present a way to obtain a perturbative solution to the invariance condition (\ref{eq:AE_inv.cond.}).
For this purpose, we expand the maps $\mathcal{K}^{(\epsilon)}$ and $\mathcal{F}^{(\epsilon)}$ with respect to $\epsilon$ as $\mathcal{K}^{(\epsilon)} = \sum_{n = 0}^\infty \epsilon^n \mathcal{K}_{n}$ and $\mathcal{F}^{(\epsilon)} = \sum_{n = 0}^\infty \epsilon^n \mathcal{F}_{n}$, respectively, where $\mathcal{K}_n$ and $\mathcal{F}_n$ are independent of $\epsilon$.
Inserting these into the invariance condition (\ref{eq:AE_inv.cond.}), the zeroth-order of $\epsilon$ reads
\begin{equation*}
    \mathcal{K}_0 \mathcal{F}_0 = \hat{\mathcal{L}}_0 \mathcal{K}_0.
\end{equation*}
Note that $\mathcal{K}_0 = \mathcal{K}^{(\epsilon = 0)}$.
From Eq. (\ref{eq:AE_K0}), we find $\mathcal{K}_0 = \chi_R$ with $\chi_R$ defined in Eq. (\ref{eq:TCL_chiRchiL}).
Using $\chi_L^\dagger \chi_R = I$, the above equation yields $\mathcal{F}^{(\epsilon = 0)} = \mathcal{F}_0 = \chi_L^\dagger \hat{\mathcal{L}}_0 \chi_R$.

For $n \geq 1$, the invariance condition at the order of $\epsilon^n$ is given by
\begin{equation}
    \mathcal{K}_0 \mathcal{F}_n + \mathcal{K}_n \mathcal{F}_0 + \sum_{m = 1}^{n-1} \mathcal{K}_m \mathcal{F}_{n-m} = \hat{\mathcal{L}}_0 \mathcal{K}_n + \hat{\mathcal{L}}_1 \mathcal{K}_{n-1}.
    \label{eq:inv.cond._order.n}
\end{equation}
This can be solved for $\mathcal{K}_n$ and $\mathcal{F}_n$ straightforwardly when $\mathcal{F}_0 = 0$ or, equivalently, when $\lambda_s = 0$ for any surviving modes $s$ \cite{FM23}.
In this case, Eq. (\ref{eq:inv.cond._order.n}) reads
\begin{equation}
    \mathcal{K}_0 \mathcal{F}_n + \sum_{m = 1}^{n-1} \mathcal{K}_m \mathcal{F}_{n-m} = \hat{\mathcal{L}}_0 \mathcal{K}_n + \hat{\mathcal{L}}_1 \mathcal{K}_{n-1}.
    \label{eq:inv.cond._order.n.F0=0}
\end{equation}
Applying $\chi_L^\dagger$ from the left yields
\begin{equation*}
    \mathcal{F}_n = \chi_L^\dagger \Big[ \hat{\mathcal{L}}_1 \mathcal{K}_{n-1} - \sum_{m = 1}^{n-1} \mathcal{K}_m \mathcal{F}_{n-m} \Big],
\end{equation*}
where we have used $\chi_L^\dagger \hat{\mathcal{L}}_0 = 0$, which is true when $\lambda_s = 0$.
Equation (\ref{eq:inv.cond._order.n.F0=0}) can be rearranged as
\begin{equation*}
    \hat{\mathcal{L}}_0 \mathcal{K}_n = \mathcal{K}_0 \mathcal{F}_n + \sum_{m = 1}^{n-1} \mathcal{K}_m \mathcal{F}_{n-m} - \hat{\mathcal{L}}_1 \mathcal{K}_{n-1}.
\end{equation*}
By inverting $\hat{\mathcal{L}}_0$, thus, we obtain $\mathcal{K}_n$.
Since $\hat{\mathcal{L}}_0$ is singular, the solution to this linear equation is not unique.
This leads to the gauge degree of freedom discussed in Sec. \ref{AE} (see Ref. \cite{TESR23} for details).

When $\lambda_s \ne 0$, Eq. (\ref{eq:inv.cond._order.n.F0=0}) similarly yields 
\begin{equation}
    \mathcal{F}_n = \chi_L^\dagger \Big[ \hat{\mathcal{L}}_0 \mathcal{K}_n + \hat{\mathcal{L}}_1 \mathcal{K}_{n-1} - \sum_{m = 1}^{n-1} \mathcal{K}_m \mathcal{F}_{n-m} \Big],
    \label{eq:inv.cond._Fn}
\end{equation}
which depends on $\mathcal{K}_n$ that has not yet been determined.
To find $\mathcal{K}_n$ and $\mathcal{F}_n$ in this case, the authors of Ref. \cite{Paolo18} guessed the form of $\mathcal{K}_n$ with an unknown quantity.
By determining the unknown quantity so that the resulting $\mathcal{K}_n$ is consistent with Eq. (\ref{eq:inv.cond._order.n}), they successfully obtained $\mathcal{K}_n$ and $\mathcal{F}_n$.
However, they calculated only up to the second order of $\epsilon$ and the analysis was limited to bipartite systems.
A systematic method for solving Eq. (\ref{eq:inv.cond._order.n}) with $\lambda_s \ne 0$ was recently developed in Ref. \cite{Angela24}.

\section{Proof of the Propositions in Sec. \ref{TCL_ad.el.}}
\label{app:proof}

In this appendix, we provide a proof of Propositions \ref{prop:relax} and \ref{prop:proj} in Sec. \ref{TCL_ad.el.}.
In what follows, we consider finite-dimensional systems and assume that $\mathcal{L}$ is diagonalizable for $0 \leq \epsilon \ll 1$ under considerations. 
This assumption is valid in various practical examples, unless the exceptional point is of interest \cite{Ashida20,Hatano19}.
Let us denote the eigenvalue problem of $\mathcal{L}$ as
(in this appendix, all superoperators are considered in their vectorized representation, and a hat is omitted for simplicity)
\begin{equation}
    (\mathcal{L} - \lambda_i^{(\epsilon)} \mathcal{I}) \dket{r_i^{(\epsilon)}} = 0, \ \ \dbra{l_i^{(\epsilon)}} (\mathcal{L} - \lambda_i^{(\epsilon)} \mathcal{I}) = 0,
    \label{eq:proof_L.eig}
\end{equation}
where the right and left eigenvectors are normalized as $\dbraket{l_i^{(\epsilon)}}{r_j^{(\epsilon)}} = \delta_{i,j}$.
The spectral decomposition of $\mathcal{L}$ yields
\begin{equation}
    \begin{gathered}
        e^{\mathcal{L} t} = \sum_i e^{\lambda_i^{(\epsilon)} t} \dketbra{r_i^{(\epsilon)}}{l_i^{(\epsilon)}} \\
        = \sum_s e^{\lambda_s^{(\epsilon)} t} \dketbra{r_s^{(\epsilon)}}{l_s^{(\epsilon)}} + \sum_f e^{\lambda_f^{(\epsilon)} t} \dketbra{r_f^{(\epsilon)}}{l_f^{(\epsilon)}}.
    \end{gathered}
    \label{eq:proof_spec.decomp.}
\end{equation}

Assuming small $\epsilon$, we identify $\lambda_i^{(\epsilon)}$, $\dket{r_i^{(\epsilon)}}$, and $\dket{l_i^{(\epsilon)}}$ as $\lim_{\epsilon \to 0} \lambda_i^{(\epsilon 
)} = \lambda_i$, $\lim_{\epsilon \to 0} \dket{r_i^{(\epsilon)}} = \dket{r_i}$, and $\lim_{\epsilon \to 0} \dbra{l_i^{(\epsilon)}} = \dbra{l_i}$, respectively.
We recall that $\lambda_i$, $\dket{r_i}$, and $\dket{l_i}$ were introduced for the eigenvalue problem of $\mathcal{L}_0$ in Eq. (\ref{eq:AE_eig.L0}).
Identifying this way, we use the subscript $s$ (surviving) and $f$ (fast relaxation) similarly for $\lambda_i^{(\epsilon)}$, $\dket{r_i^{(\epsilon)}}$, and $\dket{l_i^{(\epsilon)}}$.
In the following, we denote corrections due to the $\epsilon \mathcal{L}_1$ term as $\dket{dr_i} = \dket{r_i^{(\epsilon)}} - \dket{r_i}$ and $\dbra{dl_i} = \dbra{l_i^{(\epsilon)}} - \dbra{l_i}$.

The spectrum of $\mathcal{L}$ has a similar gap structure to that of $\mathcal{L}_0$ (see Fig. \ref{fig:AE_gap}).
The gap, denoted as $\Delta^{(\epsilon)}$, is given by 
\begin{equation*}
    \Delta^{(\epsilon)} = \min_{f,s} {\rm Re} (\lambda_f^{(\epsilon)} - \lambda_s^{(\epsilon)} ).
\end{equation*}

\subsection{Expressing $\mathcal{J}_{\rm inv}^{(\epsilon)}(t)$ and $\mathcal{P}_{\rm inv}^{(\epsilon)}(t)$ only with $\exp(\mathcal{L} t)$}

Our proof starts by recasting $\mathcal{J}_{\rm inv}^{(\epsilon)}(t)$ (Eq. (\ref{eq:TCL_J.inv.eps})) and $\mathcal{P}_{\rm inv}^{(\epsilon)}(t)$ (Eq. (\ref{eq:TCL_P.inv.eps})) into forms that involve only $\exp(\mathcal{L} t)$ but do not involve $\exp(\mathcal{Q}_{\rm inv} \mathcal{L} \mathcal{Q}_{\rm inv} t)$.
For this purpose, we first examine the definition of $\Sigma_{\rm inv}^{(\epsilon)}(t)$.
As shown in Ref. \cite{Shibata77}, the time-integral in $\Sigma_{\rm inv}^{(\epsilon)}(t)$ can be performed using
\begin{equation*}
    e^{\mathcal{Q}_{\rm inv} \mathcal{L} \mathcal{Q}_{\rm inv} t} \mathcal{Q}_{\rm inv} \mathcal{L} \mathcal{P}_{\rm inv} e^{- \mathcal{L} t}
    = - \frac{d}{dt} (e^{\mathcal{Q}_{\rm inv} \mathcal{L} \mathcal{Q}_{\rm inv} t} \mathcal{Q}_{\rm inv} e^{- \mathcal{L} t}).
\end{equation*}
Inserting this into the definition of $\Sigma_{\rm inv}^{(\epsilon)}(t)$, we obtain
\begin{equation*}
    \Sigma_{\rm inv}^{(\epsilon)}(t) = \mathcal{Q}_{\rm inv} - e^{\mathcal{Q}_{\rm inv} \mathcal{L} \mathcal{Q}_{\rm inv} t} \mathcal{Q}_{\rm inv} e^{- \mathcal{L} t}.
\end{equation*}
Using $\exp(\mathcal{Q}_{\rm inv} \mathcal{L} \mathcal{Q}_{\rm inv} t) \mathcal{P}_{\rm inv} = \mathcal{P}_{\rm inv}$, this yields
\begin{gather*}
    \mathcal{I} - \Sigma_{\rm inv}^{(\epsilon)}(t) = e^{\mathcal{Q}_{\rm inv} \mathcal{L} \mathcal{Q}_{\rm inv} t} \mathcal{Q}_{\rm inv} e^{- \mathcal{L} t} + \mathcal{P}_{\rm inv} \\
    =  e^{\mathcal{Q}_{\rm inv} \mathcal{L} \mathcal{Q}_{\rm inv} t} [\mathcal{Q}_{\rm inv} + \mathcal{P}_{\rm inv} e^{\mathcal{L} t}] e^{- \mathcal{L} t}
\end{gather*}
and 
\begin{equation*}
    [\mathcal{I} - \Sigma_{\rm inv}^{(\epsilon)}(t)]^{-1} = e^{\mathcal{L} t} [\mathcal{Q}_{\rm inv} + \mathcal{P}_{\rm inv} e^{\mathcal{L} t}]^{-1} e^{- \mathcal{Q}_{\rm inv} \mathcal{L} \mathcal{Q}_{\rm inv} t}.
\end{equation*}
Now we consider a matrix representation of $\mathcal{Q}_{\rm inv} + \mathcal{P}_{\rm inv} e^{\mathcal{L} t}$ in the basis $\{ \dket{r_i}, \dbra{l_i} \}$, namely, $[\mathcal{Q}_{\rm inv} + \mathcal{P}_{\rm inv} \exp(\mathcal{L} t)]_{ij} = \dbra{l_i} \mathcal{Q}_{\rm inv} + \mathcal{P}_{\rm inv} \exp(\mathcal{L} t) \dket{r_j}$.
Dividing into the surviving modes and the fast relaxation modes, we obtain the following block matrix representation;
\begin{equation*}
    \mathcal{Q}_{\rm inv} + \mathcal{P}_{\rm inv} e^{\mathcal{L} t} = \left[
    \begin{array}{c|c}
    A(t) & B(t) \\ \hline
    0 & I \\
    \end{array}
    \right]
    = \left[
    \begin{array}{c|c}
    I & B(t) \\ \hline
    0 & I \\
    \end{array}
    \right]
    \Pi(t)
\end{equation*}
with $A(t)_{ss'} = \dbra{l_s} \exp(\mathcal{L} t) \dket{r_{s'}}$
and $B(t)_{sf} = \dbra{l_s} \exp(\mathcal{L} t) \dket{r_f}$.
In the above equation, we have introduced $\Pi(t)$ as defined in Ref. \cite{Nestmann19};
\begin{equation*}
    \Pi(t) = \left[
    \begin{array}{c|c}
    A(t) & 0 \\ \hline
    0 & I \\
    \end{array}
    \right] = \mathcal{Q}_{\rm inv} + \mathcal{P}_{\rm inv} e^{\mathcal{L} t} \mathcal{P}_{\rm inv}.
\end{equation*}
Note that $\lim_{\epsilon \to 0} A(t)_{ss'} = \delta_{ss'} \exp(\lambda_st)$, with ${\rm Re}\lambda_s = 0$, and $[A(t)]^{-1}$ exists in the limit $\epsilon \to 0$. 
Accordingly, we expect that $[A(t)]^{-1}$ exists for small $\epsilon$.
This implies that $[\Pi(t)]^{-1}$ exists and is given by 
\begin{equation}
    \begin{gathered}
        [\Pi(t)]^{-1} = \left[
        \begin{array}{c|c}
        [A(t)]^{-1} & 0 \\ \hline
        0 & I \\
        \end{array}
        \right] \\
        = \mathcal{Q}_{\rm inv} + \mathcal{P}_{\rm inv} [\Pi(t)]^{-1} \mathcal{P}_{\rm inv}.    
    \end{gathered}
    \label{eq:proof_Pi.inv}
\end{equation}
In this case, we obtain
\begin{equation*}
\begin{gathered}
    [\mathcal{Q}_{\rm inv} + \mathcal{P}_{\rm inv} e^{\mathcal{L} t}]^{-1} = [\Pi(t)]^{-1} \left[
    \begin{array}{c|c}
    I & - B(t) \\ \hline
    0 & I \\
    \end{array}
    \right] \\
    = (\mathcal{Q}_{\rm inv} + \mathcal{P}_{\rm inv} [\Pi(t)]^{-1} \mathcal{P}_{\rm inv}) (\mathcal{I} - \mathcal{P}_{\rm inv} e^{\mathcal{L} t} \mathcal{Q}_{\rm inv}) \\
    = \mathcal{P}_{\rm inv} [\Pi(t)]^{-1} \mathcal{P}_{\rm inv} + \mathcal{Q}_{\rm inv} - \mathcal{P}_{\rm inv} [\Pi(t)]^{-1} \mathcal{P}_{\rm inv} e^{\mathcal{L} t} \mathcal{Q}_{\rm inv}.    
\end{gathered}
\end{equation*}
Therefore, $\mathcal{P}_{\rm inv}^{(\epsilon)}(t)$ and $\mathcal{J}_{\rm inv}^{(\epsilon)}(t)$ read respectively as
\begin{gather}
    \mathcal{P}_{\rm inv}^{(\epsilon)}(t) = [\mathcal{I} - \Sigma_{\rm inv}^{(\epsilon)}(t)]^{-1} \mathcal{P}_{\rm inv} \nonumber \\
    = e^{\mathcal{L} t} \mathcal{P}_{\rm inv} [\Pi(t)]^{-1} \mathcal{P}_{\rm inv},
    \label{eq:proof_P.inv.eps}
\end{gather}
and 
\begin{gather}
    \mathcal{J}_{\rm inv}^{(\epsilon)}(t) = [\mathcal{I} - \Sigma_{\rm inv}^{(\epsilon)}(t)]^{-1} e^{\mathcal{Q}_{\rm inv} \mathcal{L} \mathcal{Q}_{\rm inv} t} \mathcal{Q}_{\rm inv} \nonumber \\
    = [\mathcal{I} - \mathcal{P}_{\rm inv}^{(\epsilon)}(t)] e^{\mathcal{L} t} \mathcal{Q}_{\rm inv}.
    \label{eq:proof_J.inv.eps}
\end{gather}

\subsection{Time dependence of $\mathcal{P}_{\rm inv}^{(\epsilon)} (t)$}

Here we elucidate the time dependence of $\mathcal{P}_{\rm inv}^{(\epsilon)}(t)$ defined in Eq. (\ref{eq:proof_P.inv.eps}).
Our goal is to derive Eq. (\ref{eq:proof_P.inv.eps.final}), which decomposes $\mathcal{P}_{\rm inv}^{(\epsilon)} (t)$ into a time-independent term and the other part that decays in the limit $t \to \infty$ as shown in Appendix \ref{app:proof_projection}.
For its derivation, we introduce the two matrices $M$ and $N$ defined by $M_{s s'} = \dbraket{l_{s}^{(\epsilon)}}{r_{s'}}$ and $N_{s s'} = \dbraket{l_{s}}{r_{s'}^{(\epsilon)}}$, respectively.
In the limit $\epsilon \to 0$, they read $M_{s s'} = N_{s s'} = \delta_{s s'}$.
Given that $\epsilon$ is small, we assume the existence of $M^{-1}$ and $N^{-1}$ in the following. 

With Eq. (\ref{eq:proof_Pi.inv}), the definition of $\mathcal{P}_{\rm inv}^{(\epsilon)}(t)$ reads
\begin{equation*}
    \mathcal{P}_{\rm inv}^{(\epsilon)}(t) = \sum_{s_a, s_b} e^{\mathcal{L} t} \dket{r_{s_a}} [A(t)]_{s_a s_b}^{-1} \dbra{l_{s_b}}.
\end{equation*}
Inserting the second line of Eq. (\ref{eq:proof_spec.decomp.}) into the above expression of $\mathcal{P}_{\rm inv}^{(\epsilon)}(t)$, we obtain
\begin{equation}
    \begin{gathered}
        \mathcal{P}_{\rm inv}^{(\epsilon)}(t) 
        = \sum_{s_a, s_b, s} e^{\lambda_s^{(\epsilon)} t} \dket{r_s^{(\epsilon)}} M_{s s_a} [A(t)]_{s_a s_b}^{-1} \dbra{l_{s_b}} \\
        + \sum_{s_a, s_b} \sum_f e^{\lambda_f^{(\epsilon)} t}
        \dket{r_f^{(\epsilon)}} \dbraket{l_f^{(\epsilon)}}{r_{s_a}} [A(t)]_{s_a s_b}^{-1} \dbra{l_{s_b}}.
    \end{gathered}
    \label{eq:proof_P.inv.eps.calc}
\end{equation}

To see the time dependence of $\mathcal{P}_{\rm inv}^{(\epsilon)}(t)$ in detail, we first look into $A(t)$.
Using the spectral decomposition (\ref{eq:proof_spec.decomp.}), we find
\begin{gather*}
    A(t)_{s_a s_b} = \dbra{l_{s_a}} e^{\mathcal{L} t} \dket{r_{s_b}} \\
    = \sum_i e^{\lambda_i^{(\epsilon)} t} \dbraket{l_{s_a}}{r_{i}^{(\epsilon)}} \dbraket{l_{i}^{(\epsilon)}}{r_{s_b}} \\
    = N_{s_a s_b} e^{\lambda_{s_b}^{(\epsilon)} t} + \sum_i e^{\lambda_i^{(\epsilon)} t} \dbraket{l_{s_a}}{r_{i}^{(\epsilon)}} \dbraket{dl_{i}}{r_{s_b}},
\end{gather*}
where we have inserted $\dbra{l_i^{(\epsilon)}} = \dbra{l_i} + \dbra{dl_i}$ and used $\dbraket{l_i}{r_{s_b}} = \delta_{i s_b}$ in deriving the last line.
Let us write the last expression as $A(t) = A_1(t) + A_2(t)$ with $[A_1 (t)]_{s_a s_b} = N_{s_a s_b} e^{\lambda_{s_b}^{(\epsilon)} t}$.
The inverse of $A(t)$ can then be expressed as $[A(t)]^{-1} = [I + [A_1(t)]^{-1} A_2(t)]^{-1} [A_1(t)]^{-1}$, where 
\begin{equation*}
    [A_1(t)]_{s_a s_b}^{-1} = e^{- \lambda_{s_a}^{(\epsilon)} t} [N^{-1}]_{s_a s_b}
\end{equation*}
and 
\begin{gather}
    [[A_1(t)]^{-1} A_2(t)]_{s_a s_b} = \sum_{s'} [A_1(t)]_{s_a s'}^{-1} \ A_2(t)_{s' s_b} \nonumber \nonumber \\
    = \sum_{s'} \sum_i e^{(\lambda_{i}^{(\epsilon)} - \lambda_{s_a}^{(\epsilon)})t} [N^{-1}]_{s_a s'}  \nonumber \\
    \times \dbraket{l_{s'}}{r_{i}^{(\epsilon)}} \dbraket{dl_{i}}{r_{s_b}} \nonumber \\
    = \sum_{s,s'} e^{(\lambda_{s}^{(\epsilon)} - \lambda_{s_a}^{(\epsilon)})t} [N^{-1}]_{s_a s'} 
    N_{s' s} \dbraket{dl_{s}}{r_{s_b}} \nonumber \\
    + \sum_{s'} \sum_f e^{(\lambda_{f}^{(\epsilon)} - \lambda_{s_a}^{(\epsilon)})t} [N^{-1}]_{s_a s'} \nonumber \\
    \times \dbraket{l_{s'}}{r_{f}^{(\epsilon)}} \dbraket{dl_{f}}{r_{s_b}} \nonumber \\
    = \dbraket{dl_{s_a}}{r_{s_b}} + e^{- \Delta^{(\epsilon)} t} X(t)_{s_a s_b},
    \label{eq:proof_A1.inv.A2}
\end{gather}
where, in the last equality, we have introduced
\begin{equation}
\begin{gathered}
    X(t)_{s_a s_b} = \sum_{s'} \sum_f e^{(\lambda_{f}^{(\epsilon)} - \lambda_{s_a}^{(\epsilon)} + \Delta^{(\epsilon)}) t} \\
    \times [N^{-1}]_{s_a s'} \dbraket{l_{s'}}{r_{f}^{(\epsilon)}} \dbraket{dl_{f}}{r_{s_b}}.    
\end{gathered}
    \label{eq:proof_X}
\end{equation}
Furthermore, we find $\dbraket{dl_{s_a}}{r_{s_b}} = M_{s_a s_b} - \delta_{s_a s_b}$.
Inserting this into Eq. (\ref{eq:proof_A1.inv.A2}), we obtain
\begin{equation*}
     I + [A_1(t)]^{-1} A_2(t) = M + e^{- \Delta^{(\epsilon)} t} X(t),
\end{equation*}
which yields
\begin{equation*}
     [I + [A_1(t)]^{-1} A_2(t)]^{-1}
     = M^{-1} + e^{- \Delta^{(\epsilon)} t} M^{-1} Y(t),
\end{equation*}
with
\begin{gather}
    Y(t) = e^{\Delta^{(\epsilon)} t} \left\{ [I + e^{- \Delta^{(\epsilon)} t} X(t) M^{-1}]^{-1} - I \right\} \nonumber \\
    = - X(t)M^{-1} [I + e^{- \Delta^{(\epsilon)} t} X(t) M^{-1}]^{-1}.
    \label{eq:proof_Y}
\end{gather}
Consequently, the inverse of $A(t)$ is given by
\begin{equation*}
\begin{gathered}
    [A(t)]_{s_a s_b}^{-1} = \sum_{s'} e^{-\lambda_{s'}^{(\epsilon)} t} \\
    \times \Big( M_{s_a s'}^{-1} +  e^{- \Delta^{(\epsilon)} t} [M^{-1} Y]_{s_a s'} \Big) N_{s' s_b}^{-1} 
\end{gathered}
\end{equation*}
Inserting the above expression of $[A(t)]^{-1}$ into Eq. (\ref{eq:proof_P.inv.eps.calc}), we obtain
\begin{equation}
    \mathcal{P}_{\rm inv}^{(\epsilon)}(t) 
    = \sum_{s s'} \dket{r_{s}^{(\epsilon)}} [N^{-1}]_{s s'} \dbra{l_{s'}} + \mathcal{E}(t)
    \label{eq:proof_P.inv.eps.final}
\end{equation}
where $\mathcal{E}(t) = \mathcal{E}_1(t) + \mathcal{E}_2(t) + \mathcal{E}_3(t)$ with
\begin{gather*}
    \mathcal{E}_1(t) = \sum_{s_a, s_b, s, s'} e^{ (\lambda_s^{(\epsilon)} - \lambda_{s'}^{(\epsilon)} - \Delta^{(\epsilon)} ) t} \dket{r_{s}^{(\epsilon)}} \\
    \times Y(t)_{s s'} [N^{-1}]_{s' s_b} \dbra{l_{s_b}},
\end{gather*}
\begin{gather*}
    \mathcal{E}_2(t) = \sum_{s_a, s_b, s'} \sum_f e^{ (\lambda_f^{(\epsilon)} - \lambda_{s'}^{(\epsilon)} ) t} \dket{r_{f}^{(\epsilon)}} \\
    \times \dbraket{l_f^{(\epsilon)}}{r_{s_a}} [M]_{s_a s'}^{-1} [N]_{s' s_b}^{-1} \dbra{l_{s_b}},
\end{gather*}
and
\begin{gather*}
    \mathcal{E}_3(t) = \sum_{s_a, s_b, s'} \sum_f 
    e^{ (\lambda_f^{(\epsilon)} - \lambda_{s'}^{(\epsilon)} - \Delta^{(\epsilon)} ) t}
    \dket{r_{f}^{(\epsilon)}} \\
    \times \dbraket{l_f^{(\epsilon)}}{r_{s_a}} [M^{-1} Y(t)]_{s_a s'} [N]^{-1}_{s' s_b} \dbra{l_{s_b}}.
\end{gather*}

\subsection{Property of $\mathcal{P}_{\rm inv}^{(\epsilon)} = \lim_{t \to \infty} \mathcal{P}_{\rm inv}^{(\epsilon)} (t)$}
\label{app:proof_projection}

Here we derive the expression of $\mathcal{P}_{\rm inv}^{(\epsilon)} = \lim_{t \to \infty} \mathcal{P}_{\rm inv}^{(\epsilon)} (t)$ and discuss its properties.
By the definition of $\Delta^{(\epsilon)}$, it follows for any $(s_a,f)$ and $t \geq 0$ that 
$\left| \exp( ( \lambda_{f}^{(\epsilon)} - \lambda_{s_a}^{(\epsilon)} + \Delta^{(\epsilon)} ) t ) \right| \leq 1$.
From Eq. (\ref{eq:proof_X}), this yields
\begin{equation}
\begin{gathered}
    |X(t)_{s_a s_b}| \leq \sum_{s'} \sum_f \\
    \left| [N^{-1}]_{s_a s'} \dbraket{l_{s'}}{r_{f}^{(\epsilon)}} \dbraket{dl_{f}}{r_{s_b}} \right|,
\end{gathered}
\label{eq:proof_X.ineq}
\end{equation}
or, equivalently, $\lim_{t \to \infty} X(t)$ is a finite matrix.
From Eq. (\ref{eq:proof_Y}), this then indicates that $\lim_{t \to \infty} Y(t)$ is a finite matrix.

From these, we can show $\lim_{t \to \infty} \mathcal{E}(t) = 0$.
We focus on $\mathcal{E}_1(t)$ first.
We find for any $(s,s')$ and $t \geq 0$ that 
\begin{equation*}
    \left| e^{ (\lambda_s^{(\epsilon)} - \lambda_{s'}^{(\epsilon)} - \Delta^{(\epsilon)} ) t } \right| \leq e^{ - (\Delta^{(\epsilon)} - \gamma_{\rm slow}^{(\epsilon)} ) t }.
\end{equation*}
with $\gamma_{\rm slow}^{(\epsilon)} = \max_s {\rm Re}( - \lambda_s^{(\epsilon)} )$.
For small $\epsilon$, it follows from $\Delta^{(\epsilon)} = \Delta + O(\epsilon)$ and $\gamma_{\rm slow}^{(\epsilon)} = O(\epsilon)$ that 
\begin{equation}
    \Delta^{(\epsilon)} - \gamma_{\rm slow}^{(\epsilon)} = \Delta + O(\epsilon) > 0.
    \label{eq:proof_D-g}
\end{equation}
Given a matrix norm $\| \bullet \|$, this leads to 
\begin{gather}
\begin{gathered}
    \| \mathcal{E}_1(t) \| \ \leq e^{ - (\Delta^{(\epsilon)} - \gamma_{\rm slow}^{(\epsilon)} ) t }  \\
    \times \sum_{s_a, s_b, s, s'} \| \dket{r_{s}^{(\epsilon)}} Y(t)_{s s'} [N^{-1}]_{s' s_b} \dbra{l_{s_b}} \|
\end{gathered}      \label{eq:proof_E1}
    \\ \to 0 \ \ \ (t \to \infty).  \nonumber
\end{gather}
Similarly, we find
\begin{gather}
\begin{gathered}
    \| \mathcal{E}_2(t) \| \leq e^{- \Delta^{(\epsilon)} t} 
    \sum_{s_a, s_b, s'} \sum_f \| \dket{r_{f}^{(\epsilon)}} \\
    \times \dbraket{l_f^{(\epsilon)}}{r_{s_a}} [M]_{s_a s'}^{-1} [N]_{s' s_b}^{-1} \dbra{l_{s_b}} \|    
\end{gathered}      \label{eq:proof_E2}
    \\ \to 0  \ \ \  (t \to \infty),  \nonumber
\end{gather}
and 
\begin{gather}
\begin{gathered}
    \| \mathcal{E}_3(t) \| \leq e^{- 2 \Delta^{(\epsilon)} t} 
    \sum_{s_a, s_b, s'} \sum_f \, \| \dket{r_{f}^{(\epsilon)}} \\
     \times \dbraket{l_f^{(\epsilon)}}{r_{s_a}} [M^{-1} Y(t)]_{s_a s'} [N]^{-1}_{s' s_b} \dbra{l_{s_b}} \|     
\end{gathered}      \label{eq:proof_E3}
    \\ \to 0  \ \ \  (t \to \infty).   \nonumber
\end{gather}
Hence, $\lim_{t \to \infty} \mathcal{E}(t) = 0$ is proved.

As a result, we obtain from Eq. (\ref{eq:proof_P.inv.eps.final})
\begin{equation}
    \mathcal{P}_{\rm inv}^{(\epsilon)} = \sum_{s s'} \dket{r_{s}^{(\epsilon)}} [N^{-1}]_{s s'} \dbra{l_{s'}}.
    \label{eq:proof_P.tinf}
\end{equation}
From the definition of the matrix $N$, this equation indicates that $\mathcal{P}_{\rm inv}^{(\epsilon)}$ is a projection with $\mathcal{P}_{\rm inv}^{(\epsilon)} \dket{r_{s}^{(\epsilon)}} = \dket{r_{s}^{(\epsilon)}}$ (the image being $\mathscr{M}^{(\epsilon)}$) and $\mathcal{P}_{\rm inv}^{(\epsilon)} \dket{r_{f}} = 0$ (the kernel being a subspace spanned by the fast relaxation modes with $\epsilon = 0$).
The former property implies Eq. (\ref{eq:TCL_proj.inv.}), which can be confirmed as 
\begin{gather*}
     (\mathcal{I} - \mathcal{P}_{\rm inv}^{(\epsilon)}) \mathcal{L} \mathcal{P}_{\rm inv}^{(\epsilon)} \\
    = (\mathcal{I} - \mathcal{P}_{\rm inv}^{(\epsilon)}) \sum_{s s'} \lambda_s  \dket{r_{s}^{(\epsilon)}} [N^{-1}]_{s s'} \dbra{l_{s'}}
    = 0,
\end{gather*}
where we have used the first equation in Eqs. (\ref{eq:proof_L.eig}) in the first equality and $\mathcal{P}_{\rm inv}^{(\epsilon)} \dket{r_{s}^{(\epsilon)}} = \dket{r_{s}^{(\epsilon)}}$ in the second equality.

\subsection{Order estimate as $\epsilon \to 0$}
\label{app:proof_order}

Here we perform the order estimate of $\| \mathcal{P}_{\rm inv}^{(\epsilon)} (t) - \mathcal{P}_{\rm inv}^{(\epsilon)} \|$ and $\| \mathcal{J}_{\rm inv}^{(\epsilon)} (t) \|$ for small $\epsilon$.
To this end, we first focus on $\mathcal{E}_1 (t)$ given by Eq. (\ref{eq:proof_E1}).
Note for any $(s,f)$ that $|\dbraket{l_{s}}{r_{f}^{(\epsilon)}}| = O(\epsilon / \Delta)$ and $|\dbraket{dl_{f}}{r_{s}}| = O(\epsilon / \Delta)$.
In addition, we have $\| N \| = I + O(\epsilon)$.
From Eq. (\ref{eq:proof_X.ineq}), these lead to $\| X(t) \| = O( (\epsilon / \Delta)^2 )$.
It then follows from $\| M \| = I + O(\epsilon)$ and Eq. (\ref{eq:proof_Y}) that $\| Y(t) \| = O( (\epsilon / \Delta)^2 )$.
For the exponential time-dependent factor in Eq. (\ref{eq:proof_E1}), we note that Eq. (\ref{eq:proof_D-g}) indicates 
\begin{align*}
    |e^{ - (\Delta^{(\epsilon)} - \gamma_{\rm slow}^{(\epsilon)} ) t } | &= e^{- t \Delta (1 + O(\epsilon / \Delta)) } \\
    &= O( e^{- t \Delta})
\end{align*}
Combining these results, we obtain
\begin{equation*}
    \| \mathcal{E}_1(t) \| = O \left( (\epsilon/\Delta)^2 e^{- t \Delta} \right).
\end{equation*}
The order estimate of $\mathcal{E}_2(t)$ (Eq. (\ref{eq:proof_E2})) and $\mathcal{E}_3(t)$ (Eq. (\ref{eq:proof_E3})) can be performed similarly.
Note that $|\exp(- t \Delta^{(\epsilon)})| = \exp (- t \Delta (1 + O( \epsilon / \Delta )) ) = O (\exp(- t \Delta))$.
From $\dbraket{l_{f}^{(\epsilon)}}{r_{s}} = O(\epsilon / \Delta)$, we obtain
\begin{equation*}
    \| \mathcal{E}_2(t) \| = O \left( (\epsilon / \Delta) e^{- t \Delta} \right),
\end{equation*}
and 
\begin{equation*}
    \| \mathcal{E}_3(t) \| = O \left( (\epsilon / \Delta)^3 e^{- 2 t \Delta} \right).
\end{equation*}

For small $\epsilon$, consequently, the most dominant contribution in $\mathcal{E}(t)$ is $\mathcal{E}_2(t)$ and we obtain 
\begin{equation}
    \| \mathcal{P}_{\rm inv}^{(\epsilon)} (t) - \mathcal{P}_{\rm inv}^{(\epsilon)} \| = O \left( (\epsilon / \Delta) e^{- t \Delta} \right).
    \label{eq:proof_order.Pinv}
\end{equation}

Similarly, next we estimate the order of $\| \mathcal{J}_{\rm inv}^{(\epsilon)} (t) \|$ as $\epsilon \to 0$.
Inserting the spectral decomposition of $\exp(\mathcal{L} t)$ (Eq. (\ref{eq:proof_spec.decomp.})) and the results for $\mathcal{P}_{\rm inv}^{(\epsilon)} (t)$
(Eq. (\ref{eq:proof_P.inv.eps.final})) in Eq. (\ref{eq:proof_J.inv.eps}), we find
\begin{equation}
\begin{gathered}
    \mathcal{J}_{\rm inv}^{(\epsilon)} (t)
    = \sum_{f} e^{\lambda_f^{(\epsilon)} t} [\mathcal{I} - \mathcal{P}_{\rm inv}^{(\epsilon)}] \dket{r_f^{(\epsilon)}} \dbra{l_f^{(\epsilon)}} \mathcal{Q}_{\rm inv} \\
    - \sum_{i} e^{\lambda_i^{(\epsilon)} t} \mathcal{E}(t) \dket{r_i^{(\epsilon)}} \dbra{l_i^{(\epsilon)}} \mathcal{Q}_{\rm inv},
\end{gathered}
\label{eq:proof_J.with.E}
\end{equation}
where we have used $\mathcal{P}_{\rm inv}^{(\epsilon)} \dket{r_s^{(\epsilon)}} = \dket{r_s^{(\epsilon)}}$ in deriving this expression.
For the second line, note that $ | \exp( \lambda_f^{(\epsilon)} t ) | \leq \exp(- \Delta^{(\epsilon)} t) = O(\exp(- t \Delta))$.
From Eq. (\ref{eq:proof_P.tinf}), we find $\mathcal{I} - \mathcal{P}_{\rm inv}^{(\epsilon)} = Q_{\rm inv} + O(\epsilon)$, which then yields 
\begin{equation*}
    \| [\mathcal{I} - \mathcal{P}_{\rm inv}^{(\epsilon)}] \dket{r_f^{(\epsilon)}} \dbra{l_f^{(\epsilon)}} \mathcal{Q}_{\rm inv} \| = O(1).
\end{equation*}
Accordingly, the first term on the right hand side of Eq. (\ref{eq:proof_J.with.E}) is in the order of $O ( \exp(- t \Delta) )$.
For the second term, we can use Eq. (\ref{eq:proof_order.Pinv}) and find it to be in the order of $O ( (\epsilon / \Delta) \exp(- t \Delta)$.

For small $\epsilon$, thus, the most dominant contribution in $\mathcal{J}_{\rm inv}^{(\epsilon)}(t)$ is the first line in Eq. (\ref{eq:proof_J.with.E}) and we obtain 
\begin{equation}
    \| \mathcal{J}_{\rm inv}^{(\epsilon)} (t) \| = O (e^{- t \Delta}).
    \label{eq:proof_order.J}
\end{equation}

\section{Perturbation expansion of $\mathcal{P}_{\rm inv}^{(\epsilon)} (t)$}
\label{app:perturb}

In this appendix, we provide several formulas that are useful in a perturbation calculation of $\mathcal{P}_{\rm inv}^{(\epsilon)} (t)$.
By considering $\mathcal{P}_{\rm inv}^{(\epsilon)} = \lim_{t \to \infty} \mathcal{P}_{\rm inv}^{(\epsilon)} (t)$, one can evaluate 
$\mathcal{K}_{\rm TCL}^{(\epsilon)}$ and $\mathcal{F}_{\rm TCL}^{(\epsilon)}$ defined in Eqs. (\ref{eq:TCL_K.and.F}).
With finite $t$, the dynamics including the transient regime can be discussed (see Appendix \ref{app:Rabi_CP}).
Note that $\mathcal{J}_{\rm inv}^{(\epsilon)} (t)$ can also be evaluated from $\mathcal{P}_{\rm inv}^{(\epsilon)} (t)$ using Eq. (\ref{eq:proof_J.inv.eps}).

In what follows, we use the expressio (\ref{eq:TCL_P.inv.eps}), that is, $\mathcal{P}_{\rm inv}^{(\epsilon)}(t) 
= [\mathcal{I} - \Sigma_{\rm inv}^{(\epsilon)}(t)]^{-1} \mathcal{P}_{\rm inv}$ with
\begin{equation*}
    \Sigma_{\rm inv}^{(\epsilon)}(t) = \epsilon \int_{0}^{t} d\tau \, e^{\mathcal{Q}_{\rm inv} \mathcal{L} \mathcal{Q}_{\rm inv} \tau } \mathcal{Q}_{\rm inv} \mathcal{L}_1 \mathcal{P}_{\rm inv} e^{- \mathcal{L} \tau}.
\end{equation*}
To expand $\Sigma_{\rm inv}^{(\epsilon)} (t)$ with respect to $\epsilon$, we note the following identities;
\begin{equation*}
\begin{gathered}
    e^{(\mathcal{A}+\epsilon \mathcal{B}) (t-s)} \\
    = e^{\mathcal{A} t} \, {\rm T}_{\leftarrow} \Big\{ \exp \Big[ \epsilon \int_s^t d\tau \, \tilde{\mathcal{B}} (\tau) \Big] \Big\} \, e^{-\mathcal{A} s},
\end{gathered}
\end{equation*}
and
\begin{equation*}
    \begin{gathered}
        e^{(\mathcal{A}+\epsilon \mathcal{B}) (s-t)} \\
        = e^{\mathcal{A} s} \, {\rm T}_{\rightarrow} \Big\{ \exp \Big[ - \epsilon \int_s^t d\tau \, \tilde{\mathcal{B}} (\tau) \Big] \Big\} \, e^{-\mathcal{A} t},
    \end{gathered}
\end{equation*}
with ${\rm T}_{\leftarrow}$ and ${\rm T}_{\rightarrow}$ denoting the chronological and the antichronological time-orderings, respectively, and $\tilde{\mathcal{B}} (\tau) = \exp(- \mathcal{A} \tau) \, \mathcal{B} \, \exp(\mathcal{A} \tau)$.
These identities can be verified by considering the differential equations with respect to $t$ and $s$.
Using these, the above expression of $\Sigma_{\rm inv}^{(\epsilon)}(t)$ can be recast into the following form,
\begin{equation}
    \Sigma_{\rm inv}^{(\epsilon)}(t) = \epsilon \int_{0}^{t} d \tau \, \mathcal{G}(\tau) \mathcal{Q}_{\rm inv} \mathcal{L}_1 \mathcal{P}_{\rm inv} G(\tau),
    \label{eq:perturb_Sigma}
\end{equation}
where we have introduced ($t \geq s$),
\begin{equation*}
    \begin{gathered}
        \mathcal{G} (t-s = \tau_0 ) = e^{\mathcal{Q}_{\rm inv} \mathcal{L} \mathcal{Q}_{\rm inv} \tau_0} \\
        = e^{\mathcal{L}_0 \tau_0} \mathcal{Q}_{\rm inv} + \sum_{n = 1}^{\infty} \epsilon^n \int_0^{\tau_0} d\tau_1 \dots \int_0^{\tau_{n-1}} d\tau_n \\
        e^{\mathcal{L}_0 (\tau_0-\tau_1)} \mathcal{Q}_{\rm inv} \mathcal{L}_1 \dots e^{\mathcal{L}_0 (\tau_{n-1} - \tau_n)} \mathcal{Q}_{\rm inv} \mathcal{L}_1 e^{\mathcal{L}_0 \tau_n} \mathcal{Q}_{\rm inv},
    \end{gathered}
\end{equation*}
and
\begin{equation*}
    \begin{gathered}
        G (t-s = \tau_0 ) = e^{- \mathcal{L} \tau_0} \\
        = e^{\mathcal{L}_0 s} \, {\rm T}_{\rightarrow} \Big\{ \exp \Big[ - \epsilon \int_s^t d\tau \, \tilde{\mathcal{L}}_1 (\tau) \Big] \Big\} \, e^{- \mathcal{L}_0 t} \\
        e^{- \mathcal{L}_0 \tau_n} \mathcal{L}_1 e^{- \mathcal{L}_0 (\tau_{n-1}-\tau_n)} \dots \mathcal{L}_1 e^{- \mathcal{L}_0 (\tau_0 - \tau_1)}.
    \end{gathered}
\end{equation*}

Let us expand $\Sigma_{\rm inv}^{(\epsilon)}(t)$ as $\Sigma_{\rm inv}^{(\epsilon)}(t) = \sum_{n = 1}^\infty \epsilon^n \Sigma_n (t)$.
Inserting $\mathcal{G}(\tau)$ and $G(\tau)$ into Eq. (\ref{eq:perturb_Sigma}), we obtain, up to the order of $\epsilon^3$,
\begin{equation*}
    \Sigma_1(t) = \int_0^t d\tau e^{\mathcal{L}_0 \tau} \mathcal{Q}_{\rm inv} \mathcal{L}_1 \mathcal{P}_{\rm inv} e^{- \mathcal{L}_0 \tau},
\end{equation*}
\begin{equation}
    \begin{gathered}
        \Sigma_2(t) = \int_0^t d\tau_1 \int_0^{\tau_1} d\tau_2 \\
        \Big\{ e^{\mathcal{L}_0 (\tau_1 - \tau_2)} \mathcal{Q}_{\rm inv} \mathcal{L}_1 e^{\mathcal{L}_0 \tau_2} \mathcal{Q}_{\rm inv} \mathcal{L}_1 \mathcal{P}_{\rm inv} e^{- \mathcal{L}_0 \tau_1} \\
        - e^{\mathcal{L}_0 \tau_1} \mathcal{Q}_{\rm inv} \mathcal{L}_1 \mathcal{P}_{\rm inv} e^{- \mathcal{L}_0 \tau_2} \mathcal{L}_1 e^{- \mathcal{L}_0 (\tau_1-\tau_2)} \Big\},
    \end{gathered}
    \label{eq:perturb_Sigma2}
\end{equation}
and
\begin{equation*}
    \begin{gathered}
        \Sigma_3(t) = \int_0^t d\tau_1 \int_0^{\tau_1} d\tau_2 \int_0^{\tau_2} d\tau_3 \\
        \Big\{ e^{\mathcal{L}_0 (\tau_1 - \tau_2)} \mathcal{Q}_{\rm inv} \mathcal{L}_1 e^{\mathcal{L}_0 (\tau_2 - \tau_3)} \mathcal{Q}_{\rm inv} \mathcal{L}_1 e^{\mathcal{L}_0 \tau_3} \\ \times \mathcal{Q}_{\rm inv} \mathcal{L}_1 \mathcal{P}_{\rm inv} e^{- \mathcal{L}_0 \tau_1}
        + e^{\mathcal{L}_0 \tau_1} \mathcal{Q}_{\rm inv} \mathcal{L}_1 \mathcal{P}_{\rm inv} e^{- \mathcal{L}_0 \tau_3} \\ \times \mathcal{L}_1 e^{- \mathcal{L}_0 (\tau_2 - \tau_3)} \mathcal{L}_1 e^{- \mathcal{L}_0 (\tau_1-\tau_2)} \Big\} \\
        - \int_0^t d\tau_1 \int_0^{\tau_1} d\tau_2 \int_0^{\tau_1} d\tau_3 \,
        e^{\mathcal{L}_0 (\tau_1 - \tau_2)} \mathcal{Q}_{\rm inv} \mathcal{L}_1 e^{\mathcal{L}_0 \tau_2} \\ \times \mathcal{Q}_{\rm inv} \mathcal{L}_1 \mathcal{P}_{\rm inv} e^{- \mathcal{L}_0 \tau_3} \mathcal{L}_1 e^{- \mathcal{L}_0 (\tau_1 - \tau_3)}.
    \end{gathered}
\end{equation*}

Up to the order of $\epsilon^3$, $[\mathcal{I} - \Sigma_{\rm inv}^{(\epsilon)} (t)]^{-1}$ can be expanded as
\begin{equation*}
    \begin{gathered}
        [\mathcal{I} - \Sigma_{\rm inv}^{(\epsilon)} (t)]^{-1} = \mathcal{I} + \epsilon \Sigma_1(t) + \epsilon^2 \Sigma_2(t) \\
        + \epsilon^3 \Big[ \Sigma_3(t) + \Sigma_2(t) \Sigma_1(t) \Big] + O(\epsilon^4).
    \end{gathered}
\end{equation*}
From this, we can evaluate $\mathcal{P}_{\rm inv}^{(\epsilon)} (t)$ perturbatively. Expanding as $\mathcal{P}_{\rm inv}^{(\epsilon)} (t) = \mathcal{P}_{\rm inv} + \sum_{n = 1}^\infty \epsilon^n \mathcal{P}_n (t)$, we find
\begin{align}
    \mathcal{P}_1 (t) &=  \Sigma_1 (t) \mathcal{P}_{\rm inv}, \nonumber \\
    \mathcal{P}_2 (t) &= \Sigma_2 (t) \mathcal{P}_{\rm inv}, \label{eq:perturb_P2} \\
    \mathcal{P}_3 (t) &= \Big[ \Sigma_3(t) + \Sigma_2(t) \Sigma_1(t) \Big] \mathcal{P}_{\rm inv}, \nonumber \\
   \vdots \nonumber
\end{align}

When the limit $t \to \infty$ is taken, a simpler expression of $\mathcal{P}_{\rm inv}^{(\epsilon)}$ can be found as follows. 
Note first that $\mathcal{P}_1(t)$ reads
\begin{equation}
    \mathcal{P}_1 (t) = \int_0^t d\tau \, e^{\mathcal{L}_0 \tau} \mathcal{Q}_{\rm inv} \mathcal{L}_1 e^{- \mathcal{L}_0 \tau} \mathcal{P}_{\rm inv}.
    \label{eq:perturb_P1}
\end{equation}
Let $\mathcal{P}_n = \lim_{t \to \infty} \mathcal{P}_n (t)$.
For $\{ \mathcal{P}_n \}$, we can find relatively compact expressions by using the fact that
\begin{equation}
\begin{gathered}
    \hat{\mathcal{P}}_{\rm inv} = \sum_s \dketbra{r_s}{l_s} \equiv \sum_s \hat{\Pi}_s \ \ \ {\rm and} \\
    \hat{\mathcal{P}}_{\rm inv} = \sum_f \dketbra{r_f}{l_f} \equiv \sum_f \hat{\Pi}_f
\end{gathered}
\label{eq:perturb_PQ}
\end{equation}
are projections onto the eigenspaces of $\mathcal{L}_0$.
For instance, $\mathcal{P}_1$ reads from Eq. (\ref{eq:perturb_P1}) as 
\begin{equation}
\begin{gathered}
    \mathcal{P}_1 = \int_0^\infty d\tau \, e^{\mathcal{L}_0 \tau} \mathcal{Q}_{\rm inv} \mathcal{L}_1 \mathcal{P}_{\rm inv} e^{- \mathcal{L}_0 \tau} \mathcal{P}_{\rm inv} \nonumber\\
    = \sum_{sf} \left( \int_0^\infty d\tau e^{(\lambda_f - \lambda_s) \tau} \right) \Pi_f \mathcal{L}_1 \Pi_s \nonumber\\
    = \sum_{s f} \, \frac{1}{\Delta_{sf}} \, \Pi_f \mathcal{L}_1 \Pi_s,
\end{gathered}\label{eq:projector_P1}
\end{equation}
where we have introduced $\Delta_{sf} = \lambda_s - \lambda_f$, which has the positive real part as ${\rm Re}\Delta_{sf} \geq \Delta > 0$.
Similar calculations yield higher-order contributions as 
\begin{equation}
    \mathcal{P}_{\rm inv}^{(\epsilon)} = \mathcal{P}_{\rm inv} + \sum_{sf} \, \frac{1}{\Delta_{sf}} \Pi_f \Big[ \epsilon \mathcal{L}_1 + \sum_{n = 2}^\infty \epsilon^n \Gamma_{n,fs} \Big] \Pi_s,
    \label{eq:perturb_Pn}
\end{equation}
with 
\begin{equation}
    \Gamma_{2,fs} = \mathcal{L}_1 \frac{\mathcal{Q}_{\rm inv}}{\lambda_s - \mathcal{L}_0} \mathcal{L}_1 - \mathcal{L}_1 \frac{\mathcal{P}_{\rm inv}}{\mathcal{L}_0 - \lambda_f} \mathcal{L}_1,
    \label{eq:perturb_Gamma2}
\end{equation}
and 
\begin{equation}
    \begin{gathered}
        \Gamma_{3,fs} =
        \mathcal{L}_1 \frac{\mathcal{Q}_{\rm inv}}{\lambda_s - \mathcal{L}_0} \mathcal{L}_1 \frac{\mathcal{Q}_{\rm inv}}{\lambda_s - \mathcal{L}_0} \mathcal{L}_1 \\
        + \mathcal{L}_1 \frac{\mathcal{P}_{\rm inv}}{\mathcal{L}_0 - \lambda_f} \mathcal{L}_1 \frac{\mathcal{P}_{\rm inv}}{\mathcal{L}_0 - \lambda_f} \mathcal{L}_1 \\
        - \mathcal{L}_1 \frac{\mathcal{Q}_{\rm inv}}{\lambda_s - \mathcal{L}_0} \mathcal{L}_1 \frac{\mathcal{P}_{\rm inv}}{\mathcal{L}_0 - \lambda_f} \mathcal{L}_1 \\
        - \mathcal{L}_1 \frac{\mathcal{P}_{\rm inv}}{\mathcal{L}_0 - \lambda_f} \mathcal{L}_1 \frac{\mathcal{Q}_{\rm inv}}{\lambda_s - \mathcal{L}_0} \mathcal{L}_1 \\
        - \sum_{s' f'} \, \frac{\Delta_{sf}}{\Delta_{s' f'}} \, \mathcal{L}_1 \frac{\Pi_{f'}}{\Delta_{s f'}} \mathcal{L}_1 \frac{\Pi_{s'}}{\Delta_{s' f}} \mathcal{L}_1.
    \end{gathered}
    \label{eq:perturb_Gamma3}
\end{equation}

\section{Comparison with the Laplace transform method}
\label{app:Laplace}

For bipartite systems, a formulation of adiabatic elimination based on the projection superoperator was considered in Ref. \cite{Saideh20}.
In this appendix, we show that the generator introduced in Ref. \cite{Saideh20} is in general inconsistent with the one in the TCL formulation (and thus the geometric formulation).
The following calculations adopt the general definition of $\mathcal{P}$ given by Eq. (\ref{eq:TCL_P.def}), which we here denote as $\hat{\mathcal{P}} = \sum_s \dketbra{r_s}{l_s}$  without the subscript "inv" to conform to their notation.
For bipartite systems, this definition is consistent with the one in Ref. \cite{Saideh20} as we demonstrate in Eq. (\ref{eq:projection_operation}).
In what follows, we use the notations $\mathcal{P} = \sum_s \Pi_s$ and $\mathcal{Q} = \sum_f \Pi_f$ as defined in Eq. (\ref{eq:perturb_PQ}).

Instead of the time-convolutionless master equation, the authors of Ref. \cite{Saideh20} focused on the Nakajima-Zwanzig equation (\ref{eq:TCL_NZ}).
Assuming $\mathcal{Q}\rho(0) = 0$, Eq. (\ref{eq:TCL_NZ}) reads
\begin{equation*}
    \begin{gathered}
        \frac{d}{dt} \mathcal{P} \rho(t) = \mathcal{P} \mathcal{L} \mathcal{P} \rho(t) \\
        + \int_{0}^t ds \, \mathcal{P} \mathcal{L} \mathcal{Q} e^{\mathcal{Q} \mathcal{L} \mathcal{Q} (t-s) } \mathcal{Q} \mathcal{L} \mathcal{P} \rho(s).
    \end{gathered}
\end{equation*}
While this equation is nonlocal with respect to the time argument, we can obtain a local equation by applying the Laplace transform.
The effective generator in the Laplace domain, $\mathcal{L}_{\rm eff} (z)$, is given by 
\begin{equation*}
    \mathcal{L}_{\rm eff} (z) = \mathcal{P} \mathcal{L} \mathcal{P} + \mathcal{P} \mathcal{L} \mathcal{Q} [z \mathcal{I} - \mathcal{Q} \mathcal{L} \mathcal{Q}]^{-1} \mathcal{Q} \mathcal{L} \mathcal{P}.
\end{equation*}
The authors of Ref. \cite{Saideh20} argued that the dynamics in a long-time domain can be described by $\mathcal{L}_{\rm eff} (z)$ near $z = 0$.
By expanding $\mathcal{L}_{\rm eff} (z)$ around $z = 0$ and executing the inverse Laplace transform, they found the evolution equation
\begin{equation*}
    \frac{d}{dt} \mathcal{P}\rho(t) = [\mathcal{I} - \mathcal{M}_1]^{-1} \mathcal{M}_0 \mathcal{P}\rho(t),
\end{equation*}
with 
\begin{equation*}
    \begin{gathered}
        \mathcal{M}_0 = \mathcal{L}_{\rm eff} (z=0) \\
        = \mathcal{P} \mathcal{L} \mathcal{P} - \mathcal{P} \mathcal{L} \mathcal{Q} [ \mathcal{Q} \mathcal{L} \mathcal{Q}]^{-1} \mathcal{Q} \mathcal{L} \mathcal{P}
    \end{gathered}
\end{equation*}
and 
\begin{equation*}
    \begin{gathered}
        \mathcal{M}_1 = \left. \frac{d}{dz} \mathcal{L}_{\rm eff} (z) \right\vert_{z = 0}  \\
        = - \mathcal{P} \mathcal{L} \mathcal{Q} [ \mathcal{Q} \mathcal{L} \mathcal{Q}]^{-2} \mathcal{Q} \mathcal{L} \mathcal{P}.
    \end{gathered}
\end{equation*}
In these equations, $[ \mathcal{Q} \mathcal{L} \mathcal{Q}]^{-1}$ is defined as the inverse of $\mathcal{L}$ in the subspace spanned by $\{ \Pi_f \}$.
For instance, $\mathcal{Q} \mathcal{L}_0 \mathcal{Q} = \sum_f \lambda_f \Pi_f$ and $[ \mathcal{Q} \mathcal{L}_0 \mathcal{Q}]^{-1} = \sum_f \lambda_f^{-1} \Pi_f = \mathcal{Q}/\mathcal{L}_0$.

We evaluate the generator $[\mathcal{I} - \mathcal{M}_1]^{-1} \mathcal{M}_0$ approximately using the perturbation expansion with respect to $\epsilon$.
Let us first consider the expansion up to the second order of $\epsilon$.
Note that
\begin{equation*}
    \mathcal{M}_0 = \mathcal{P} (\mathcal{L}_0 + \epsilon \mathcal{L}_1) \mathcal{P} - \epsilon^2 \mathcal{P} \mathcal{L}_1 \mathcal{Q} [ \mathcal{Q} \mathcal{L} \mathcal{Q}]^{-1} \mathcal{Q} \mathcal{L}_1 \mathcal{P},
\end{equation*}
and 
\begin{equation*}
    \mathcal{M}_1 = - \epsilon^2 \mathcal{P} \mathcal{L}_1 \mathcal{Q} [ \mathcal{Q} \mathcal{L} \mathcal{Q}]^{-2} \mathcal{Q} \mathcal{L}_1 \mathcal{P}.
\end{equation*}
Accordingly, we have
\begin{equation*}
    \begin{gathered}
       [\mathcal{I} - \mathcal{M}_1]^{-1} \\
       = \mathcal{I} - \epsilon^2 \mathcal{P} \mathcal{L}_1 \mathcal{Q} [ \mathcal{Q} \mathcal{L} \mathcal{Q}]^{-2} \mathcal{Q} \mathcal{L}_1 \mathcal{P} + O(\epsilon^4).
    \end{gathered}
\end{equation*}
and 
\begin{equation*}
    \begin{gathered}
       [\mathcal{I} - \mathcal{M}_1]^{-1} \mathcal{M}_0 = \mathcal{P} (\mathcal{L}_0 + \epsilon \mathcal{L}_1) \mathcal{P} \\
       - \epsilon^2 \mathcal{P} \mathcal{L}_1 \Big\{ [ \mathcal{Q} \mathcal{L} \mathcal{Q}]^{-2} \mathcal{Q} \mathcal{L}_1 \mathcal{L}_0 + [ \mathcal{Q} \mathcal{L} \mathcal{Q}]^{-1} \mathcal{Q} \mathcal{L}_1 \Big\}\mathcal{P} \\
       + O(\epsilon^3).
    \end{gathered}
\end{equation*}
Inserting $[ \mathcal{Q} \mathcal{L} \mathcal{Q}]^{-n} = \sum_f \lambda_f^{-n} \Pi_f + O(\epsilon) \ (n = 1,2)$ into this equation, we obtain
\begin{equation}
    \begin{gathered}
       [\mathcal{I} - \mathcal{M}_1]^{-1} \mathcal{M}_0 = \mathcal{P} (\mathcal{L}_0 + \epsilon \mathcal{L}_1) \mathcal{P} \\
       + \epsilon^2 \mathcal{P} \mathcal{L}_1 \sum_{sf} \frac{1}{(-\lambda_f)} \Pi_f \Big( \frac{\lambda_s + \lambda_f}{\lambda_f} \mathcal{L}_1  \Big) \Pi_s \\
       + O(\epsilon^3).
    \end{gathered}
    \label{eq:Laplace_2nd.Saideh}
\end{equation}
On the other hand, the evolution of $\mathcal{P} \rho(t)$ in the time-convolutionless master equation is given by $\mathcal{P} \mathcal{L} \mathcal{P}^{(\epsilon)}$.
With Eq. (\ref{eq:perturb_Pn}), we find 
\begin{equation}
    \begin{gathered}
        \mathcal{P} \mathcal{L} \mathcal{P}^{(\epsilon)} = \mathcal{P} (\mathcal{L}_0 + \epsilon \mathcal{L}_1) \mathcal{P} \\
       + \epsilon^2 \mathcal{P} \mathcal{L}_1 \sum_{sf} \frac{1}{(\lambda_s-\lambda_f)} \Pi_f \mathcal{L}_1 \Pi_s \\
       + O(\epsilon^3).
    \end{gathered}
    \label{eq:Laplace_2nd.TCL}
\end{equation}
At the second order of $\epsilon$, thus, the generator introduced in Ref. \cite{Saideh20} [Eq. (\ref{eq:Laplace_2nd.Saideh})] is in general inconsistent with the one in the time-convolutionless master equation [Eq. (\ref{eq:Laplace_2nd.TCL})].
This finding indicates that the former is inconsistent with the geometric formulation. 
Below Eq. (34) in Ref. \cite{Saideh20}, it was argued that the second-order expansion of their proposed generator agrees with the result obtained from the geometric formulation for a bipartite system.
On this point, it should be noted that Eqs. (\ref{eq:Laplace_2nd.Saideh}) and (\ref{eq:Laplace_2nd.TCL}) agree if $\lambda_s = 0$.
In fact, the systems considered in Ref. \cite{Saideh20} have the property $\lambda_s = 0$, and this explains why the agreement with the geometric formulation was observed.

Now a question emerges regarding the consistency between $[\mathcal{I} - \mathcal{M}_1]^{-1} \mathcal{M}_0$ and $\mathcal{P} \mathcal{L} \mathcal{P}^{(\epsilon)}$ at higher orders of $\epsilon$, under the assumption of $\lambda_s = 0$.
To address this question, we first perform the perturbation expansion of $[ \mathcal{Q} \mathcal{L} \mathcal{Q}]^{-n} \ (n = 1,2)$.
Using the formula $[\mathcal{A} - \epsilon \mathcal{B}]^{-1} = \mathcal{A}^{-1} + \epsilon \mathcal{A}^{-1} \mathcal{B} \mathcal{A}^{-1} + \epsilon^2 \mathcal{A}^{-1} \mathcal{B} \mathcal{A}^{-1} \mathcal{B} \mathcal{A}^{-1} + O(\epsilon^3)$ for linear maps $\mathcal{A}$ and $\mathcal{B}$, we find
\begin{equation}
    \begin{gathered}
        [\mathcal{Q} \mathcal{L} \mathcal{Q}]^{-1} = \sum_f \frac{1}{\lambda_f} \Pi_f - \epsilon \sum_f \frac{1}{\lambda_f} \Pi_f \mathcal{L}_1 \frac{\mathcal{Q}}{\mathcal{L}_0} \\
        + \epsilon^2 \sum_{f} \frac{1}{\lambda_f} \Pi_f \mathcal{L}_1 \frac{\mathcal{Q}}{\mathcal{L}_0} \mathcal{L}_1 \frac{\mathcal{Q}}{\mathcal{L}_0}
        + O(\epsilon^3),
    \end{gathered}
    \label{eq:Laplace_inv.1}
\end{equation}
which leads to
\begin{equation}
    \begin{gathered}
        [\mathcal{Q} \mathcal{L} \mathcal{Q}]^{-2} = \sum_f \frac{1}{\lambda_f^2} \Pi_f \\
        - \epsilon \sum_f \frac{1}{\lambda_f} \Big[ \frac{1}{\lambda_f} \Pi_f \mathcal{L}_1 \frac{\mathcal{Q}}{\mathcal{L}_0} + \Pi_f \mathcal{L}_1 \frac{\mathcal{Q}}{\mathcal{L}_0^2} \Big] + O(\epsilon^2).
    \end{gathered}
    \label{eq:Laplace_inv.2}
\end{equation}
Since $\mathcal{L}_0 \mathcal{P} = 0$ when $\lambda_s = 0$, we have
\begin{equation*}
    \begin{gathered}
       [\mathcal{I} - \mathcal{M}_1]^{-1} \mathcal{M}_0 \\
       = [\mathcal{I} - \epsilon^2 \mathcal{P} \mathcal{L}_1 \mathcal{Q} [ \mathcal{Q} \mathcal{L} \mathcal{Q}]^{-2} \mathcal{Q} \mathcal{L}_1 \mathcal{P} + O(\epsilon^4)] \\
       \times [\epsilon \mathcal{P} \mathcal{L}_1 \mathcal{P} - \epsilon^2 \mathcal{P} \mathcal{L}_1 \mathcal{Q} [ \mathcal{Q} \mathcal{L} \mathcal{Q}]^{-1} \mathcal{Q} \mathcal{L}_1 \mathcal{P}] \\
       = \epsilon \mathcal{P} \mathcal{L}_1 \mathcal{P} -  \epsilon^2 \mathcal{P} \mathcal{L}_1 \mathcal{Q} [ \mathcal{Q} \mathcal{L} \mathcal{Q}]^{-1} \mathcal{Q} \mathcal{L}_1 \mathcal{P} \\
       - \epsilon^3 \mathcal{P} \mathcal{L}_1 \mathcal{Q} [ \mathcal{Q} \mathcal{L} \mathcal{Q}]^{-2} \mathcal{Q} \mathcal{L}_1 \mathcal{P} \mathcal{L}_1 \mathcal{P} \\
       + \epsilon^4 \mathcal{P} \mathcal{L}_1 \mathcal{Q} [ \mathcal{Q} \mathcal{L} \mathcal{Q}]^{-2} \mathcal{Q} \mathcal{L}_1 \mathcal{P} \mathcal{L}_1 \mathcal{Q} [ \mathcal{Q} \mathcal{L} \mathcal{Q}]^{-1} \mathcal{Q} \mathcal{L}_1 \mathcal{P} \\
       + O(\epsilon^5).
    \end{gathered}
\end{equation*}
Inserting Eqs. (\ref{eq:Laplace_inv.1}) and (\ref{eq:Laplace_inv.2}), we obtain the third-order contribution 
\begin{equation*}
    \begin{gathered}
        \mathcal{P} \mathcal{L}_1 \sum_f \frac{1}{\lambda_f} \Pi_f \mathcal{L}_1 \frac{\mathcal{Q}}{\mathcal{L}_0} \mathcal{L}_1 \mathcal{P}
        - \mathcal{P} \mathcal{L}_1 \mathcal{Q} \sum_{f} \frac{1}{\lambda_f^2} \Pi_f  \mathcal{L}_1 \mathcal{P} \mathcal{L}_1 \mathcal{P} \\
        = \sum_{sf} \frac{1}{(-\lambda_f)} \mathcal{P} \mathcal{L}_1 \Pi_f \Big\{ - \mathcal{L}_1 \frac{\mathcal{Q}}{\mathcal{L}_0} \mathcal{L}_1 + \mathcal{L}_1 \frac{\mathcal{P}}{\lambda_f} \mathcal{L}_1 \Big\} \Pi_s,
    \end{gathered}
\end{equation*}
and the fourth-order contribution 
\begin{equation*}
    \begin{gathered}
        \mathcal{P} \mathcal{L}_1 \sum_{f} \frac{1}{\lambda_f} \Pi_f \mathcal{L}_1 \frac{\mathcal{Q}}{\mathcal{L}_0} \mathcal{L}_1 \frac{\mathcal{Q}}{\mathcal{L}_0} \mathcal{L}_1 \mathcal{P} \\
        + \mathcal{P} \mathcal{L}_1 \sum_f \frac{1}{\lambda_f} \left( \frac{1}{\lambda_f} \Pi_f \mathcal{L}_1 \frac{\mathcal{Q}}{\mathcal{L}_0} + \Pi_f \mathcal{L}_1 \frac{\mathcal{Q}}{\mathcal{L}_0^2} \right) \mathcal{L}_1 \mathcal{P} \mathcal{L}_1 \mathcal{P} \\
        + \mathcal{P} \mathcal{L}_1 \sum_f \frac{1}{\lambda_f^2} \Pi_f  \mathcal{L}_1 \mathcal{P} \mathcal{L}_1 \frac{\mathcal{Q}}{\mathcal{L}_0} s\mathcal{L}_1 \mathcal{P} \\
        = \sum_{sf} \frac{1}{(-\lambda_f)} \mathcal{P} \mathcal{L}_1 \Pi_f \left( \mathcal{L}_1 \frac{\mathcal{Q}}{\mathcal{L}_0} \mathcal{L}_1 \frac{\mathcal{Q}}{\mathcal{L}_0} \mathcal{L}_1 - \mathcal{L}_1 \frac{\mathcal{Q}}{\mathcal{L}_0} \mathcal{L}_1 \frac{\mathcal{P}}{\lambda_f} \mathcal{L}_1 \right. \\
        \left. - \mathcal{L}_1 \frac{\mathcal{Q}}{\mathcal{L}_0^2} \mathcal{L}_1 \mathcal{P} \mathcal{L}_1 - \mathcal{L}_1 \frac{\mathcal{P}}{\lambda_f} \mathcal{L}_1 \frac{\mathcal{Q}}{\mathcal{L}_0} \mathcal{L}_1 \right) \Pi_s.
    \end{gathered}
\end{equation*}
To see the consistency with $\mathcal{P} \mathcal{L} \mathcal{P}^{(\epsilon)}$, we only need to compare the terms inside $\{ \}$ in the above equations with $\Gamma_{n,fs} \ (n = 2,3)$ in Eq. (\ref{eq:perturb_Pn}) with $\lambda_s = 0$.
From Eq. (\ref{eq:perturb_Gamma2}), the third-order contributions agree with each other.
On the other hand, from Eq. (\ref{eq:perturb_Gamma3}), the fourth-order contributions do not agree due to the lack of the term $\mathcal{L}_1 (\mathcal{P}/\lambda_f) \mathcal{L}_1 (\mathcal{P}/\lambda_f) \mathcal{L}_1$ in $[\mathcal{I} - \mathcal{M}_1]^{-1} \mathcal{M}_0$.
Consequently, even under the assumption of $\lambda_s = 0$, the generator introduced in Ref. \cite{Saideh20} does not agree with the one in the time-convolutionless master equation (and thus, the geometric formulation) at the fourth-order of $\epsilon$.

\section{Numerical analysis of the equivalence for the Rabi model}
\label{app:infdim}

In Sec. \ref{Demo_bipartite}, we apply the TCL master equation to the Rabi model, which involves a photon system with an infinite-dimensional space.
The results in Appendix \ref{app:proof}, summarized in Propositions \ref{prop:relax} and \ref{prop:proj}, do not directly extend to infinite-dimensional systems.
In this appendix, we present numerical studies to support the equivalence for the Rabi model.

The Rabi model is defined by the generator $\mathcal{L} = \mathcal{L}_0 + \epsilon \mathcal{L}_1$, where $\mathcal{L}_0 = \mathcal{L}_A \otimes \mathcal{I}_B + \mathcal{I}_A \otimes \mathcal{L}_B$ and $\epsilon \mathcal{L}_1 = \mathcal{L}_{\rm int}$. The components $\mathcal{L}_A$, $\mathcal{L}_B$, and $\mathcal{L}_{\rm int}$ are respectively specified in Eqs. (\ref{eq:Demo_Rabi.LA}), (\ref{eq:Demo_Rabi.LB}), and (\ref{eq:Demo_Rabi.Lint}).
For numerical computations, we use the Fock basis $\{ \ket{n} \}_{n \in \mathbb{Z}_{\geq 0}}$ to represent photon operators and truncate the space to dimension $n_{\rm tr}$ by setting $\mel{m}{O_A}{n} = 0$ whenever $m \geq n_{\rm tr}$ or $n \geq n_{\rm tr}$ for any photon operator $O_A$. The parameters are chosen as $\omega_{\rm ph} = \omega_{\rm eg} = \kappa$ and $g = 0.05 \kappa$.

The propositions in Sec. \ref{TCL_ad.el.} are inapplicable to infinite-dimensional systems.
To illustrate this, we use the Rabi model to analyze the dependence of $\| d \hat{\mathcal{P}}_{\rm inv}^{(\epsilon)}(t) \|_F$ and $\| \hat{\mathcal{J}}_{\rm inv}^{(\epsilon)}(t) \|_F$ (defined in Sec. \ref{Demo_prop}) on the truncation dimension $n_{\rm tr}$.
These results fit well to a single exponential function,
$\| d \hat{\mathcal{P}}_{\rm inv}^{(\epsilon)}(t) \|_F = a_P \exp(- b_P t \Delta)$ and $\| \hat{\mathcal{J}}_{\rm inv}^{(\epsilon)}(t) \|_F = a_J \exp(- b_J t \Delta)$ with the gap $\Delta = \kappa/2$ (see Appendix \ref{app:Rabi}), as in Fig. \ref{fig:proof_3level} (b).
Numerical fitting for $\kappa t \in [5,30]$ yields coefficients listed in Table \ref{tab:infdim_fitparams}. While the decay rate parameters ($b_P$ and $b_J$) are nearly constant and close to unity, the amplitude parameters ($a_P$ and $a_J$) grow with $n_{\rm tr}$.
This behavior can be understood by considering the case $\epsilon = 0$.
From Eq. (\ref{eq:TCL_J.inv.eps}) and $\Sigma_{\rm inv}^{(\epsilon = 0)}(t) = 0$, we find
\begin{equation*}
    \hat{\mathcal{J}}_{\rm inv}^{(\epsilon = 0)}(t) = \sum_f e^{\lambda_f t} \dketbra{r_f}{l_f}.
\end{equation*}
Although this quantity decays exponentially, its matrix norm diverges in infinite-dimensional systems.
Consequently, we expect $a_P, a_J \to \infty$ in the limit $n_{\rm tr} \to \infty$.

\begin{table}[h]
  \centering
  \begin{tabular}{|c|c|c|c|c|} \hline
    $n_{\rm tr}$ & $a_P$ & $a_J$ & $b_P$ & $b_J$ \\ \hline
    10 & 0.655 & 19.1 & 0.98193 & 1.0066 \\ \hline
    20 & 0.926 & 39.1 & 0.98193 & 1.0063 \\ \hline
    30 & 1.13 & 59.0 & 0.98193 & 1.0061 \\ \hline
  \end{tabular}
  \caption{
  Dependence of the fitting parameters $a_P$, $a_J$, $b_P$, and $b_J$ ($\| d \hat{\mathcal{P}}_{\rm inv}^{(\epsilon)}(t) \|_F = a_P e^{- b_P t \Delta}$ and $\| \hat{\mathcal{J}}_{\rm inv}^{(\epsilon)}(t) \|_F = a_J e^{- b_J t \Delta}$) on the truncation dimension $n_{\rm tr}$.
  }
  \label{tab:infdim_fitparams}
\end{table}

To circumvent this divergence in the analysis, we focus on the density operator rather than the superoperators, as the goal of adiabatic elimination is to obtain the state evolution. 
Equation (\ref{eq:TCL_full.rho}), $\dket{\rho(t)} = \hat{\mathcal{P}}_{\rm inv}^{(\epsilon)}(t) \dket{\rho(t)}$, can be expressed as
\begin{equation*}
\begin{gathered}
    \dket{\rho(t)} - \left( \sum_{s s'} \dket{r_{s}^{(\epsilon)}} [N^{-1}]_{s s'} \dbra{l_{s'}} \right) \dket{\rho(t)} \\
    = \hat{\mathcal{J}}_{\rm inv}^{(\epsilon)}(t) \dket{\mathcal{Q}_{\rm inv} \rho(0)} + d \hat{\mathcal{P}}_{\rm inv}^{(\epsilon)}(t) \dket{\rho(t)},
\end{gathered}
\end{equation*}
where the right-hand side represents the error term $e_\mathcal{K}(t)$ in Eqs. (\ref{eq:TCL_full.x.error}).
As discussed, our objective is to verify that this error term becomes negligible in the long-time regime.
Thus, instead of the norms of the superoperators $\hat{\mathcal{J}}_{\rm inv}^{(\epsilon)}(t)$ and $d \hat{\mathcal{P}}_{\rm inv}^{(\epsilon)}(t)$, we focus on the norms of the operators $\hat{\mathcal{J}}_{\rm inv}^{(\epsilon)}(t) \dket{\mathcal{Q}_{\rm inv} \rho(0)}$ and $d \hat{\mathcal{P}}_{\rm inv}^{(\epsilon)}(t) \dket{\rho(t)}$.

To simplify the analysis, we assume $\mathcal{Q}_{\rm inv} \rho(0) = 0$. 
In the bipartite case, this choice corresponds to an initial product state including the unique steady state of $\mathcal{L}_A$ ($\ketbra{0}{0}$ in the Rabi model, see Appendix \ref{app:Rabi}).
The error term then reduces to $d \hat{\mathcal{P}}_{\rm inv}^{(\epsilon)}(t) \dket{\rho(t)}$, and its norm is the focus of our analysis.
Assuming an initial state $\rho(0) = \ketbra{0}{0} \otimes \ketbra{e}{e}$, we find that $n_{\rm tr} = 10$ suffices for density operator computation under the current settings.
Convergence is confirmed by comparing density operators computed at $n_{\rm tr}$ and $n_{\rm tr} + 5$, showing discrepancies below $10^{-13}$ for $\kappa t \in [0,50]$ when $n_{\rm tr} \geq 10$.
Additionally, we verify $\| \dket{\rho(t)} -\hat{\mathcal{P}}_{\rm inv}^{(\epsilon)}(t) \dket{\rho(t)} \|_F \sim 10^{-16}$ (machine precision) in the same time region.

\begin{figure}[t]
  \includegraphics[keepaspectratio, scale=0.5]{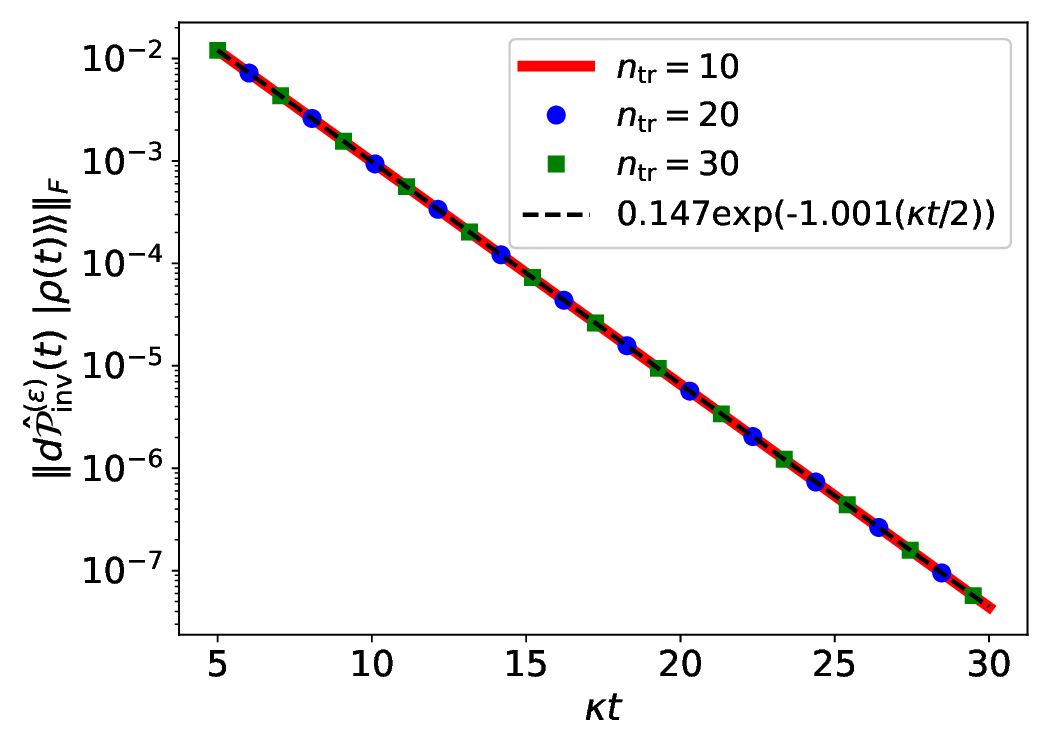}
  \caption{
  Exponential decay of the Frobenius norms $\| d \hat{\mathcal{P}}_{\rm inv}^{(\epsilon)}(t) \dket{\rho(t)} \|_F$ at truncation dimensions $n_{\rm tr} = 10$ (thick solid red line), $20$ (blue circles), and $30$ (green squares). The thin dashed black line is the single-exponential fit applied to all the truncation dimensions.
  }
  \label{fig:infdim_Rabi}
\end{figure}

Figure \ref{fig:infdim_Rabi} shows the $n_{\rm tr}$-dependence of the norm $\| d \hat{\mathcal{P}}_{\rm inv}^{(\epsilon)}(t) \dket{\rho(t)} \|_F$.
Unlike $\| d \hat{\mathcal{P}}_{\rm inv}^{(\epsilon)}(t) \|_F$, this norm is nearly independent of the truncation dimension $n_{\rm tr}$.
The results for different $n_{\rm tr}$ values fit well to a single exponential function, $\| d \hat{\mathcal{P}}_{\rm inv}^{(\epsilon)}(t) \dket{\rho(t)} \|_F = a \exp(- b t \Delta)$ with $a = 0.147$ and $1.001$, for $\kappa t \in [5,30]$.
In contrast to $a_P$ in Table \ref{tab:infdim_fitparams}, the amplitude parameter, $a$, is nearly independent of $n_{\rm tr}$.
This finding suggests that the error term vanishes in the long-time regime irrespective of the truncation dimension, thereby supporting the equivalence for the Rabi model.

\section{Adiabatic elimination for the Rabi model including a lossy photon mode}
\label{app:Rabi}

In this appendix, we present details of the second-order adiabatic elimination calculation for the Rabi model considered in Sec. \ref{Demo_bipartite}.
The generator reads $\mathcal{L} = \mathcal{L}_0 + \epsilon \mathcal{L}_1$ with $\mathcal{L}_0 = \mathcal{L}_A \otimes \mathcal{I}_B + \mathcal{I}_A \otimes \mathcal{L}_B$ and $\epsilon \mathcal{L}_1 = \mathcal{L}_{\rm int}$. The generators $\mathcal{L}_A$, $\mathcal{L}_B$, and $\mathcal{L}_{\rm int}$ are respectively defined in Eqs. (\ref{eq:Demo_Rabi.LA}), (\ref{eq:Demo_Rabi.LB}), and (\ref{eq:Demo_Rabi.Lint}).
In this case, the eigenvalue problem of $\mathcal{L}_A$ can be solved analytically as (see Appendix C of Ref. \cite{TESR23})
\begin{equation*}
    (\hat{\mathcal{L}}_A - \lambda_{A,mn}\hat{\mathcal{I}}) \dket{r_{A,mn}} = 0, \ \ \dbra{l_{A,mn}} (\hat{\mathcal{L}}_A - \lambda_{A,mn}\hat{\mathcal{I}}) = 0,
\end{equation*}
for $m,n \in \mathbb{Z}_{\geq 0}$, where the eigenvalues are given by $\lambda_{A,mn} = - (\bar{\kappa}m + \bar{\kappa}^* n) / 2$, with $\bar{\kappa} = \kappa + 2 i \omega_{\rm ph}$, and the right and left eigenvectors are given by $\dket{r_{A,mn}} = \exp(- \hat{\mathcal{A}}) \dket{ \ketbra{m}{n} }$, $\dbra{l_{A,mn}} = \dbra{ \ketbra{m}{n} } \exp(\hat{\mathcal{A}})$, respectively, with $\{ \ket{n} \}_{n \in \mathbb{Z}_{\geq 0}}$ the Fock states of the photon mode and $\mathcal{A} \rho = a \rho a^\dagger$.
The mode $m = n = 0$ is the only mode with the eigenvalue of 0.
Hence, the steady state $\bar{\rho}_A$ is unique and is given by
\begin{equation*}
    \bar{\rho}_A = r_{A,00} = \ketbra{0}{0}.
\end{equation*}

As shown in Eq. (\ref{eq:projection_vectorized}), the projection $\mathcal{P}_{\rm inv}$ in the current example reads 
\begin{gather*}
    \hat{\mathcal{P}}_{\rm inv} = \dketbra{\bar{\rho}_A}{I_A} \otimes \hat{\mathcal{I}}_B \\
    = \dketbra{\ketbra{0}{0}}{I_A} \otimes \hat{\mathcal{I}}_B.
\end{gather*}
From this, we can show 
\begin{equation}
    \mathcal{P}_{\rm inv} \mathcal{L}_1 \mathcal{P}_{\rm inv} = 0,
    \label{eq:Rabi_P.L1.P}
\end{equation}
as $\mathcal{L}_1$ is in the odd power of $a$ and $a^\dagger$.
In addition, we find the expression of $\mathcal{Q}_{\rm inv} = \mathcal{I} - \mathcal{P}_{\rm inv}$ as
\begin{equation*}
    \hat{\mathcal{Q}}_{\rm inv} = \sum_{\substack{m,n = 0 \\ m+n > 0}}^\infty 
  \dketbra{r_{A,mn}}{l_{A,mn}} \otimes \hat{\mathcal{I}}_B.
\end{equation*}
Since $\dket{r_{A,mn}}$ are the right eigenvectors of $\mathcal{L}_A$, this yields, for an arbitrary qubit operator $X$,
\begin{equation}
    \begin{gathered}
        e^{\mathcal{L}_0 t} \mathcal{Q}_{\rm inv} (\ketbra{m}{n} \otimes X + (H.c.)) \\
        = \sum_{\substack{k,q = 0 \\ k+q > 0}}^\infty e^{-(\bar{\kappa} k + \bar{\kappa}^* q)t/2} \mel{k}{e^{\mathcal{A}} (\ketbra{m}{n}) }{q}  \\
        \times e^{- \mathcal{A}} (\ketbra{k}{q}) \otimes e^{\mathcal{L}_B t} ( X ) + (H.c.),
    \end{gathered}
    \label{eq:Rabi_eL.Qinv}
\end{equation}
where $H.c.$ denotes the Hermitian conjugate of the preceding terms.

Our goal is to evaluate $\mathcal{P}_{\rm inv}^{(\epsilon)} (t)$ up to the second-order of $\epsilon$.
The perturbation expansion of $\mathcal{P}_{\rm inv}^{(\epsilon)} (t)$, $\mathcal{P}_{\rm inv}^{(\epsilon)} (t) = \mathcal{P}_{\rm inv} + \sum_{n=1}^\infty \epsilon^n \mathcal{P}_n(t)$, is performed in Appendix \ref{app:perturb}.
The second-order contribution, $\mathcal{P}_2 (t)$, is given by Eq. (\ref{eq:perturb_P2}) with $\Sigma_2 (t)$ defined in Eq. (\ref{eq:perturb_Sigma2}).
Utilizing Eq. (\ref{eq:Rabi_P.L1.P}), it reads
\begin{equation}
    \begin{gathered}
        \epsilon^2 \mathcal{P}_2 (t) \rho = \epsilon^2 \int_0^t d\tau_1 \int_0^{\tau_1} d\tau_2 \, e^{\mathcal{L}_0 (\tau_1 - \tau_2)} \\
        \times \mathcal{Q}_{\rm inv} \mathcal{L}_1 e^{\mathcal{L}_0 \tau_2} \mathcal{Q}_{\rm inv} \mathcal{L}_1 \mathcal{P}_{\rm inv} e^{- \mathcal{L}_0 \tau_1} \rho.
    \end{gathered}
    \label{eq:Rabi_P2}
\end{equation}
The expression of the integrand can be simplified using Eq. (\ref{eq:Rabi_eL.Qinv}).
For later use, we calculate it sequentially as
\begin{equation}
\begin{gathered}
    \epsilon e^{\mathcal{L}_0 \tau_2} \mathcal{Q}_{\rm inv} \mathcal{L}_1 \mathcal{P}_{\rm inv} e^{- \mathcal{L}_0 \tau_1} \rho \\
    = - i g e^{- \bar{\kappa} \tau_2 / 2} \ketbra{1}{0} \otimes e^{\mathcal{L}_B \tau_2} ( \sigma_x e^{- \mathcal{L}_B \tau_1} (\rho_B) ) \\
    + (H.c.),
\end{gathered}
\label{eq:Rabi_for.P1}
\end{equation}
and
\begin{equation}
\begin{gathered}
    \epsilon^2 e^{\mathcal{L}_0 (\tau_1 - \tau_2)} \mathcal{Q}_{\rm inv} \mathcal{L}_1 e^{\mathcal{L}_0 \tau_2} \mathcal{Q}_{\rm inv} \mathcal{L}_1 \mathcal{P}_{\rm inv} e^{- \mathcal{L}_0 \tau_1} \rho \\
    = g^2  e^{- \bar{\kappa} \tau_2 / 2} \\
    \times \Big\{ e^{- (\bar{\kappa} + \bar{\kappa}^*) (\tau_1 - \tau_2) / 2} (\ketbra{1}{1} - \ketbra{0}{0}) \\
    \otimes e^{\mathcal{L}_B (\tau_1 - \tau_2)} (e^{\mathcal{L}_B \tau_2} ( \sigma_x e^{- \mathcal{L}_B \tau_1} (\rho_B) ) \sigma_x) \\ 
    + \sqrt{2} e^{- \bar{\kappa} (\tau_1 - \tau_2)} \ketbra{2}{0} \\ \otimes e^{\mathcal{L}_B (\tau_1 - \tau_2)} (\sigma_x e^{\mathcal{L}_B \tau_2} ( \sigma_x e^{- \mathcal{L}_B \tau_1} (\rho_B) )) \Big\} \\
    + (H.c.).
\end{gathered}
\label{eq:Rabi_for.P2}
\end{equation}

It follows from $\mathcal{L}_B \rho = - i (\omega_{\rm eg} / 2) [\sigma_z, \rho]$ that the operation of $\exp(\mathcal{L}_B t)$ is given by
\begin{equation*}
    e^{\mathcal{L}_B t} (\rho) = U_B(t) \rho U_B^\dagger(t),
\end{equation*}
with $U_B(t) = \exp(- i (\omega_{\rm eg} t / 2) \sigma_z)$ being the unitary operator $U_B^\dagger(t) = U_B(- t) = U_B(t)^{-1}$.
By unitary transformation, $\sigma_x$ is transformed as
\begin{equation*}
    U_B(t) \sigma_x U_B^\dagger(t) = \sigma_- e^{i \omega_{\rm eg} t} + \sigma_+ e^{- i \omega_{\rm eg} t}.
\end{equation*}
Using these relations, we can evaluate the parts involving $\rho_B$ in Eqs. (\ref{eq:Rabi_for.P1}) and (\ref{eq:Rabi_for.P2}).

\subsection{Second-order expansion of $\mathcal{F}_{\rm TCL}^{(\epsilon)}$ and $\mathcal{K}_{\rm TCL}^{(\epsilon)}$}

We now evaluate the maps $\mathcal{F}_{\rm TCL}^{(\epsilon)}$ and $\mathcal{K}_{\rm TCL}^{(\epsilon)}$ up to the second-order of $\epsilon$.
As a parametrization, we consider $\rho_B = {\rm tr}_A (\rho)$ [see Eq. (\ref{eq:Demo_rhoB}) for its validity].
It then follows that the operations of $\chi_R$ and $\chi_L^\dagger$ defined in Eq. (\ref{eq:TCL_chiRchiL}) are given by $\chi_R \rho_B = \ketbra{0}{0} \otimes \rho_B$ and $\chi_L^\dagger \rho = {\rm tr}_A (\rho)$, respectively.
Inserting these into the definition of the maps $\mathcal{F}_{\rm TCL}^{(\epsilon)}$ and $\mathcal{K}_{\rm TCL}^{(\epsilon)}$, Eqs. (\ref{eq:TCL_K.and.F}), we find, up to the second-order of $\epsilon$, 
\begin{gather}
    \mathcal{F}_{\rm TCL}^{(\epsilon)} \rho_B = {\rm tr}_A (\mathcal{L} \mathcal{P}_{\rm inv}^{(\epsilon)} \rho) \nonumber \\
    = \mathcal{L}_B \rho_B + \epsilon^2 {\rm tr}_A (\mathcal{L}_1 \mathcal{P}_1 \rho ),
    \label{eq:Rabi_FTCL.expand}
\end{gather}
and 
\begin{gather}
    \mathcal{K}_{\rm TCL}^{(\epsilon)} \rho_B = \mathcal{P}_{\rm inv}^{(\epsilon)} \rho \nonumber \\
    = (\mathcal{P}_{\rm inv} + \epsilon \mathcal{P}_1 + \epsilon^2 \mathcal{P}_2) \rho,
    \label{eq:Rabi_KTCL.expand}
\end{gather}
with $\mathcal{P}_n = \lim_{t \to \infty} \mathcal{P}_n (t) \ (n = 1,2)$. Note that we have used Eq. (\ref{eq:Rabi_P.L1.P}) to derive Eq. (\ref{eq:Rabi_FTCL.expand}).

The first-order contribution $\mathcal{P}_1(t)$ is given by Eq. (\ref{eq:perturb_P1}), the integrand of which is given by Eq. (\ref{eq:Rabi_for.P1}) with $\tau_1 = \tau_2 = \tau$.
Accordingly, we obtain
\begin{equation}
    \epsilon \mathcal{P}_1(t) = -i g \ketbra{0}{0} \otimes \sigma_\gamma (t) \rho_B + (H.c.),
    \label{eq:Rabi_P1}
\end{equation}
where we have introduced
\begin{equation*}
    \sigma_\gamma (t) = c_-(t) \sigma_- + c_+(t) \sigma_+,
\end{equation*}
with $c_\pm (t) = [ 1 - \exp(- \gamma_\pm t) ] / \gamma_\pm$ and $\gamma_\pm = \bar{\kappa}/2 \pm i \omega_{\rm eg}$.
This then leads to 
\begin{gather}
    \epsilon^2 {\rm tr}_A (\mathcal{L}_1 \mathcal{P}_1(t) \rho) \nonumber \\
    = - g^2 \Big\{ \sigma_x \sigma_\gamma (t) \rho_B
    - \sigma_\gamma (t) \rho_B \sigma_x \Big\} + (H.c.) \nonumber \\
    \begin{gathered}
        = - \frac{i g^2}{2} {\rm Im}(c_+(t) - c_-(t) ) [\sigma_z, \rho_B] \\ 
        + g^2 \sum_{j,k = \pm} K_{jk}(t) \Bigg[ \sigma_j \rho_B \sigma_k^\dagger -\frac{\sigma_k^\dagger \sigma_j \rho_B + \rho_B \sigma_k^\dagger \sigma_j}{2} \Bigg],
    \end{gathered}
    \label{eq:Rabi_FTCL2.td}
\end{gather}
with $K_{jk}(t) = c_j(t) + c_k^*(t)$.
We thus obtain from Eq. (\ref{eq:Rabi_FTCL.expand}) the second-order reduced dynamics $(d/dt) \rho_B (t) = \mathcal{F}_{\rm TCL}^{(\epsilon)} \rho_B (t)$ with 
\begin{equation}
\begin{gathered}
    \mathcal{F}_{\rm TCL}^{(\epsilon)} \rho_B = - \frac{i}{2} \Big\{ \omega_{\rm eg} + g^2 {\rm Im}(c_+ - c_-) \Big\} [\sigma_z, \rho_B] \\ 
    + g^2 \sum_{j,k = \pm} K_{jk} \Bigg[\sigma_j \rho_B \sigma_k^\dagger -\frac{\sigma_k^\dagger \sigma_j \rho_B + \rho_B \sigma_k^\dagger \sigma_j}{2} \Bigg],
\end{gathered}
\label{eq:Rabi_FTCL}
\end{equation}
where $c_\pm = \lim_{t \to \infty} c_\pm (t) = 1/\gamma_\pm $ and $K_{jk} = \lim_{t \to \infty} K_{jk}(t) = c_j + c_k^*$. 

The second-order contribution $\mathcal{P}_2$ can be evaluated by inserting Eq. (\ref{eq:Rabi_for.P2}) into Eq. (\ref{eq:Rabi_P2}). As a result, we obtain
\begin{gather*}
    \epsilon^2 \mathcal{P}_2 \rho = \mathcal{X} \rho_B \\
    - g^2 \Bigg[ \sqrt{2} \ketbra{2}{0} \otimes \Big( \frac{\sigma_- \sigma_+}{\bar{\kappa} \gamma_+} + \frac{\sigma_+ \sigma_-}{\bar{\kappa} \gamma_-} \Big) \rho_B + (H.c.) \Bigg], \\
\end{gather*}
with
\begin{gather*}
    \mathcal{X} \rho_B = g^2 (\ketbra{1}{1} - \ketbra{0}{0}) \\
    \otimes \Big( \frac{\sigma_- \rho_B \sigma_-}{\gamma_- (\gamma_- + \gamma_+^*)} 
    + \frac{\sigma_+ \rho_B \sigma_+}{\gamma_+ (\gamma_+ + \gamma_-^*)} \\ + \frac{\sigma_- \rho_B \sigma_+}{\gamma_- (\gamma_- + \gamma_-^*)} + \frac{\sigma_+ \rho_B \sigma_-}{\gamma_+ (\gamma_+ + \gamma_+^*)} \Big) + (H.c.) \\
    = g^2 (\ketbra{1}{1} - \ketbra{0}{0}) \otimes \sigma_\gamma \rho_B \sigma_\gamma^\dagger,
\end{gather*}
with $\sigma_\gamma = \lim_{t \to \infty} \sigma_\gamma(t) = c_- \sigma_- + c_+ \sigma_+$. 
Inserting this and Eq. (\ref{eq:Rabi_P1}) in the limit $t \to \infty$ into Eq. (\ref{eq:Rabi_KTCL.expand}), we find 
\begin{gather*}
    \mathcal{K}_{\rm TCL}^{(\epsilon)} \rho_B = \ketbra{0}{0} \otimes \rho_B \\
    + W (\ketbra{0}{0} \otimes \rho_B)  + (\ketbra{0}{0} \otimes \rho_B) W^\dagger + \mathcal{X} \rho_B,
\end{gather*}
with
\begin{gather*}
    W = - ig a^\dagger \otimes \sigma_\gamma
    - g^2 (a^\dagger)^2 \otimes \Big( \frac{\sigma_- \sigma_+}{\bar{\kappa} \gamma_+} + \frac{\sigma_+ \sigma_-}{\bar{\kappa} \gamma_-} \Big).
\end{gather*}
Note that
\begin{gather*}
    \mathcal{K}_{\rm TCL}^{(\epsilon)} \rho_B = (I + W) (\ketbra{0}{0} \otimes \rho_B) (I + W)^\dagger \\
    \mathcal{X} \rho_B - W (\ketbra{0}{0} \otimes \rho_B) W^\dagger, 
\end{gather*}
where $I$ is the identity operator on the total space and
\begin{gather*}
    \mathcal{X} \rho_B - W (\ketbra{0}{0} \otimes \rho_B) W^\dagger 
    = - g^2 \ketbra{0}{0} \otimes \sigma_\gamma \rho_B \sigma_\gamma^\dagger \\
    + O(g^3).
\end{gather*}
Therefore, within the accuracy of the second-order expansion, we obtain $\rho(t) = \mathcal{K}_{\rm TCL}^{(\epsilon)} \rho_B (t)$ with
\begin{equation}
\begin{gathered}
    \mathcal{K}_{\rm TCL}^{(\epsilon)} \rho_B = (I + W) (\ketbra{0}{0} \otimes \rho_B) (I + W)^\dagger \\
    - g^2 (I_A \otimes \sigma_\gamma) (\ketbra{0}{0} \otimes \rho_B) (I_A \otimes \sigma_\gamma)^\dagger.
\end{gathered}
\label{eq:Rabi_KTCL}
\end{equation}

\begin{figure}[t]
  \includegraphics[keepaspectratio, scale=0.38]{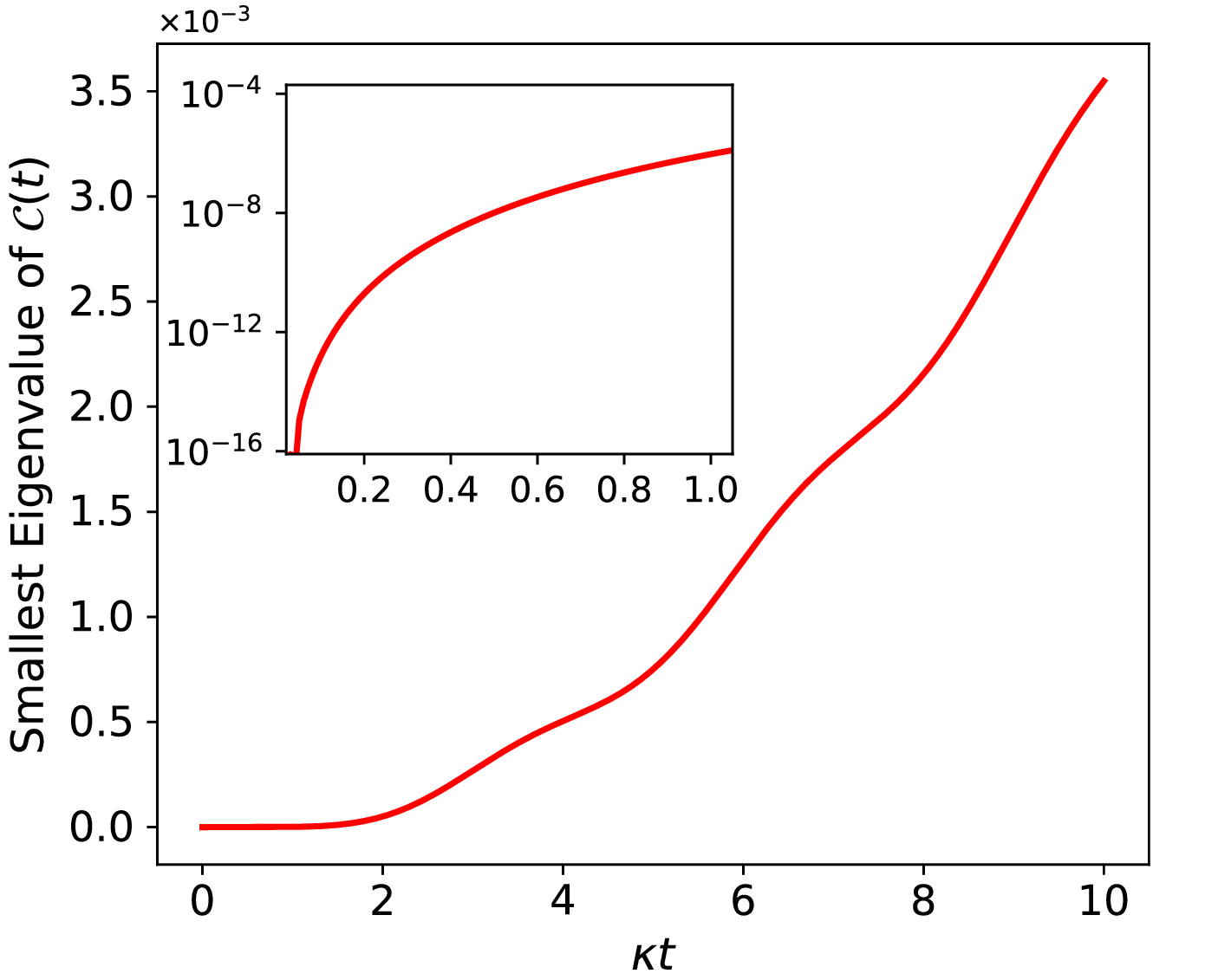}
  \caption{
  Nonnegativity of the smallest eigenvalue of the Choi matrix $\mathcal{C}(t)$ [see Eq. (\ref{eq:Rabi_Choi})].
  The inset is a zoom in on the region $0 \leq \kappa t \leq 1.05$.
  }
  \label{fig:Rabi_Choi}
\end{figure}

\subsection{Recovering complete positivity by incorporating the time-dependence}
\label{app:Rabi_CP}

As discussed with Eq. (\ref{eq:Demo_K.eigval}), the generator $\mathcal{F}_{\rm TCL}^{(\epsilon)}$ given by Eq. (\ref{eq:Rabi_FTCL}) is not in the GKSL form and the evolution is not completely positive. 
Here we numerically show that the evolution generated by the time-dependent counterpart $\mathcal{F}_{\rm TCL}^{(\epsilon)} (t)$ defined by Eq. (\ref{eq:Demo_FTCL.td}) is completely positive.

As the second-order contribution is given by Eq. (\ref{eq:Rabi_FTCL2.td}), $\mathcal{F}_{\rm TCL}^{(\epsilon)} (t)$ can be obtained by replacing $c_\pm$ and $K_{jk}$ in Eq. (\ref{eq:Rabi_FTCL}) by $c_\pm (t)$ and $K_{jk} (t)$, respectively. 
Since $[\mathcal{F}_{\rm TCL}^{(\epsilon)} (t_1), \mathcal{F}_{\rm TCL}^{(\epsilon)} (t_2)] \ne 0 \ (t_1 \ne t_2)$ in general, the propagator is given with the chronological time-ordering $T_{\leftarrow}$ as 
\begin{equation*}
    T_{\leftarrow} \Big\{ e^{ \int_0^t d \tau \mathcal{F}_{\rm TCL}^{(\epsilon)} (\tau)} \Big\}.
\end{equation*}

To confirm its complete positivity, we consider the Choi matrix representation
\begin{equation}
    \mathcal{C}(t) = \sum_{p,q = e,g} \ketbra{p}{q} \otimes T_{\leftarrow} \Big\{ e^{ \int_0^t d \tau \mathcal{F}_{\rm TCL}^{(\epsilon)} (\tau)} \Big\} (\ketbra{p}{q}),
    \label{eq:Rabi_Choi}
\end{equation}
which is positive semidefinite if and only if the propagator is completely positive \cite{Choi75}.
The Choi matrix $\mathcal{C}(t)$ can be evaluated by solving  $(d/dt) \rho_B (t) = \mathcal{F}_{\rm TCL}^{(\epsilon)} (t) \rho_B (t)$ with different initial states $\rho_B(t = 0) = \ketbra{p}{q} \ (p,q = e,g)$.
For numerical computations, we arbitrarily set $\omega_{\rm ph} = \omega_{\rm eg} = \kappa$ and $g = 0.1 \kappa$.
The differential equation was solved using the fourth-order Runge-Kutta method with the time step size $10^{-3} \kappa^{-1}$.

Figure \ref{fig:Rabi_Choi} shows the smallest eigenvalue of $\mathcal{C}(t)$ as a function of time.
Except for the infinitesimal time region, the smallest eigenvalue is positive. 
In the infinitesimal time region, note that
\begin{equation*}
    K (dt) = 2 dt
    \begin{pmatrix}
        1 & 1 \\
        1 & 1
    \end{pmatrix}
    + O(dt^2).
\end{equation*}
Accordingly, the matrix $K$ is positive semidefinite to the leading order, ensuring complete positivity of the infinitesimal evolution.
In conclusion, the propagator is completely positive at all times.


\end{document}